\documentclass[acmsmall,nonacm]{acmart}

\usepackage{multirow}
\usepackage{nomencl}
\usepackage{dirtytalk}
\usepackage{amsmath}

\usepackage{amssymb}
\usepackage{amsthm}
\usepackage{mathtools}
\usepackage{subfig}

\usepackage{hyperref}
\usepackage[capitalize,noabbrev]{cleveref} %
\usepackage{url}
\usepackage{microtype}
\usepackage{relsize}
\usepackage{soul}

\usepackage{algorithm}
\usepackage{algorithmicx}
\usepackage{algpseudocode}
\usepackage{arydshln}

\usepackage{listings}
\usepackage{xspace}
\usepackage{multirow}
\usepackage{thmtools}
\usepackage{thm-restate}
\usepackage{breqn}

\usepackage{booktabs}       %
\usepackage{tabularx}
\usepackage{nicefrac}       %
\usepackage{microtype}      %
\usepackage{colortbl}       %
\usepackage{stackengine}
\usepackage{lipsum}
\usepackage{dsfont}
\usepackage{graphicx}
\usepackage{tabularx}
\usepackage{float}
\usepackage{mathtools}
\usepackage{mathrsfs}
\usepackage{enumitem}
\usepackage{xspace}
\usepackage{multicol}

\usepackage{xparse}
\usepackage[export]{adjustbox}

\usepackage{conf/tikzit}

\tikzstyle{BLACK}=[draw=black, shape=circle, fill=black, inner sep=3pt]
\tikzstyle{DOM}=[fill={rgb,255: red,122; green,0; blue,42}, draw=black, shape=circle, inner sep=3pt]
\tikzstyle{NONE}=[fill={rgb,255: red,247; green,33; blue,212}, draw=black, shape=circle, inner sep=3pt]
\tikzstyle{NTWO}=[fill={rgb,255: red,0; green,177; blue,219}, draw=black, shape=circle, inner sep=3pt]
\tikzstyle{NTHR}=[fill={rgb,255: red,140; green,143; blue,14}, draw=black, shape=circle, inner sep=3pt]
\tikzstyle{GRAYN}=[fill={rgb,255: red,128; green,128; blue,128}, draw=black, shape=circle, inner sep=3pt]
\tikzstyle{BRGRAY}=[fill={rgb,255: red,236; green,236; blue,236}, draw=none, shape=circle, inner sep=3pt]
\tikzstyle{TEXTSTD}=[fill=white, draw=black, shape=rectangle, tikzit shape=rectangle, text width=3cm, rounded corners]
\tikzstyle{TEXTHIGH}=[fill={rgb,255: red,231; green,245; blue,255}, draw=black, shape=rectangle, tikzit shape=rectangle, text width=3cm, rounded corners]
\tikzstyle{TEXTHIGHGRAY}=[fill={rgb,255: red,220; green,220; blue,220}, draw=black, shape=rectangle, tikzit shape=rectangle, text width=3cm, rounded corners]
\tikzstyle{TEXTHIGHGREEN}=[fill={rgb,255: red,149; green,242; blue,145}, draw=black, shape=rectangle, tikzit shape=rectangle, text width=3.4cm, rounded corners]
\tikzstyle{TEXTHIGHYELLOW}=[fill={rgb,255: red,250; green,250; blue,160}, draw=black, shape=rectangle, tikzit shape=rectangle, text width=3cm, rounded corners]

\tikzstyle{EDGE}=[-, fill=none]
\tikzstyle{BLUE}=[-, draw={rgb,255: red,0; green,101; blue,189}]
\tikzstyle{GRAY}=[-, fill={rgb,255: red,128; green,128; blue,128}]
\tikzstyle{DARKGREEN}=[-, fill=none, draw={rgb,255: red,19; green,100; blue,13}, line width=1.25pt]
\tikzstyle{big dash}=[-, dashed, dash pattern=on 4mm off 2mm, fill={rgb,255: red,178; green,255; blue,253}]
\tikzstyle{big dash thick}=[-, thick, dashed, dash pattern=on 4mm off 2mm, fill={rgb,255: red,178; green,255; blue,253}]
\tikzstyle{new edge style 0}=[-, fill={rgb,255: red,224; green,254; blue,255}]
\tikzstyle{FILLGREEN}=[-, fill={rgb,255: red,232; green,255; blue,207}]
\tikzstyle{FILLBLUE}=[-, fill={rgb,255: red,231; green,255; blue,255}]
\tikzstyle{FILLPURPLE}=[-, fill={rgb,255: red,255; green,233; blue,239}]
\tikzstyle{FILLDARKBLUE}=[-, fill={rgb,255: red,238; green,230; blue,255}]
\tikzstyle{FILLDARKBLUEINVISIBLE}=[-, draw=none, fill={rgb,255: red,238; green,230; blue,255}]
\tikzstyle{FILLGREENINVISBLE}=[-, draw=none, fill={rgb,255: red,232; green,255; blue,207}]
\tikzstyle{ULTRAGRAY}=[-, draw={rgb,255: red,150; green,150; blue,150}]
\tikzstyle{EDGEDASHED}=[-, dashed, fill=none]
\tikzstyle{simple directed edge}=[->, draw=black, thick, line width=0.75mm]
\tikzstyle{simple dashed directed edge}=[->, draw=black, dashed]
\tikzstyle{SLIGHTLYDASHED}=[-, dotted, fill=none, draw={rgb,255: red,128; green,128; blue,128}]
\tikzstyle{simple dashed directed edge}=[->, draw=black, line width=0.75mm, dashed]
\tikzstyle{simple dashed directed edge not thick}=[->, draw=black, line width=0.45mm, dashed]
\tikzstyle{simple}=[->, draw=black, thick]
\tikzstyle{simple}=[draw=black, fill=none, tikzit draw=black, ->]
\tikzstyle{invis}=[-, draw=none]
\tikzstyle{red-edge}=[-, draw={rgb,255: red,191; green,0; blue,64}, fill=none, line width=1.5pt]

\usepackage[framemethod=TikZ]{mdframed}
\usepackage{footnote}
\usepackage{booktabs}
\usepackage{soul}

    \definecolor{orangeish}{HTML}{f1a340} %
    \definecolor{grayish}{HTML}{f7f7f7} %
    \definecolor{purpleish}{HTML}{998ec3} %
    \definecolor{blueish}{HTML}{004488}
    \definecolor{darkgreen}{RGB}{2,100,64} %
    \definecolor{lightgreen}{HTML}{b2f2bb}
    \definecolor{lightblueish}{RGB}{230,244,255}
\PassOptionsToPackage{table,xcdraw,svgnames,dvipsnames}{xcolor}

\newtheorem{theorem}{Theorem}[section]
\newtheorem{lemma}[theorem]{Lemma}

\newtheorem{example}[theorem]{Example}
\newtheorem{definition}[theorem]{Definition}

\newtheorem{corollary}[theorem]{Corollary}
\newtheorem{claim}[theorem]{Claim}
\newtheorem{fact}[theorem]{Fact}

\renewenvironment{quote}
  {\list{}{\rightmargin=0.3cm \leftmargin=0.3cm}%
   \item\relax}
  {\endlist}

\AtBeginDocument{%
  }

\renewcommand{\epsilon}{\varepsilon}

\newcommand{\privateMSTFramework}{\ensuremath{\mathcal{A}
    _{\textit{\scalebox{0.7}{\textit{priv-MST}}}}}\xspace}

\newcommand{\privateKruskal}{\ensuremath{\mathcal{A}
    _{\textit{\scalebox{0.7}{\textit{priv-kruskal}}}}}\xspace}

\newcommand{\onePassPrivateKruskal}{\ensuremath{\mathcal{A}
    _{\textit{\scalebox{0.7}{\textit{one-pass-priv-kruskal}}}}}\xspace}

\DeclareCaptionFormat{custom}
{%
    \textbf{#1#2}\textit{\small #3}
}
\newcommand{\R}{\ensuremath{\mathbb{R}}}
\newcommand{\N}{\ensuremath{\mathbb{N}}}

\crefname{lstlisting}{Listing}{Listings}

\renewcommand\vec{\mathbf}

\newcommand{\cA}{\mathcal{A}}
\newcommand{\cB}{\mathcal{B}}
\newcommand{\cC}{\mathcal{C}}

\newcommand{\cE}{\mathcal{E}}

\newcommand{\cG}{\mathcal{G}}
\newcommand{\cI}{\mathcal{I}}
\newcommand{\cJ}{\mathcal{J}}

\newcommand{\cM}{\mathcal{M}}

\newcommand{\cT}{\mathcal{T}}

\newcommand{\cX}{\mathcal{X}}
\newcommand{\cY}{\mathcal{Y}}

\DeclareMathOperator*{\argmin}{arg\,min}
\newcommand{\norm}[1]{\left\Vert {#1} \right\Vert}
\newcommand{\PAREN}[1]{{\left( {#1} \right)}}

\newcommand{\Bigparen}[1]{{\Big( {#1} \Big)}}
\newcommand{\bigparen}[1]{{\big( {#1} \big)}}
\newcommand{\paren}[1]{{( {#1} )}}

\newcommand{\card}[1]{\left| {#1} \right|}
\newcommand{\set}[1]{\left\{ {#1} \right\}}
\newcommand{\eps}{\varepsilon}

\NewDocumentCommand\p{ m g }{
  \ensuremath{
    \IfNoValueTF{#2}
    {\Pr [ #1 ]}
    {\Pr_{#1}[#2]}
  }
}
\RenewDocumentCommand\P{ m g }{
  \ensuremath{
    \IfNoValueTF{#2}
    {\Pr \left[#1\right]}
    {\Pr_{#1}\left[#2\right]}
  }
}

\newcommand{\BetaFunt}[2]{\mathrm{B}( {#1}, {#2})}
\newcommand{\BetaDist}[2]{\operatorname{\mathcal{B}eta}\left( {#1}, {#2} \right)}

\DeclareMathOperator*{\Expectation}{\mathbb{E}}
\NewDocumentCommand\E{ m g }{
  \ensuremath{
    \IfNoValueTF{#2}
    {\Expectation \left[#1\right]}
    {\Expectation_{#1}\left[#2\right]}
  }
}
\DeclareMathOperator*{\Variance}{\mathbb{V}\mathrm{ar}}
\NewDocumentCommand\Var{ m g }{
  \ensuremath{
    \IfNoValueTF{#2}
    {\Variance \left[#1\right]}
    {\Variance_{#1}\left[#2\right]}
  }
}

\NewDocumentCommand\GammaDist{ m g }{
  \ensuremath{
    \IfNoValueTF{#2}
    {\operatorname{\mathcal{G}amma}\left( {#1} \right)}
    {\operatorname{\mathcal{G}amma}\left( {#1}, {#2} \right)}
  }
}

\NewDocumentCommand\LapNoise{ m g }{
  \ensuremath{
    \IfNoValueTF{#2}
    {\operatorname{\mathbb{L}ap}\left( {#1} \right)}
    {\operatorname{\mathbb{L}ap}\left( {#1}, {#2} \right)}
  }
}

\NewDocumentCommand\GumbelNoise{ m g }{
  \ensuremath{
    \IfNoValueTF{#2}
    {\operatorname{\mathbb{G}umbel}\left( {#1} \right)}
    {\operatorname{\mathbb{G}umbel}\left( {#1}, {#2} \right)}
  }
}

\newcommand{\PPSACR}{{\it PPSACR}\xspace}
\newcommand{\OneShotPPSACR}{{\it One-Shot-PPSACR}\xspace}
\newcommand{\numSample}{s}
\newcommand{\SpanningTrees}[1]{\cT \PAREN{#1}}

\newcommand{\NameExpoMesm}{\textsc{exp}}
\newcommand{\ExpoMesm}{\mathcal{M}_{\textsc{exp}}}

\newcommand{\SensitivityExpoMesm}{\Delta_{\textsc{exp}}}

\newcommand{\IntSet}[2]{[{#1}\,.\,.\,{#2}]}

\NewDocumentCommand\loss{g g g}{
  \ensuremath{
    {\IfNoValueTF{#3}
        { 
            \IfNoValueTF{#2} 
                {
                    \IfNoValueTF{#1}
                        {\cE}
                        {\cE_{#1}}
                }
                {\cE\paren{{#1, #2}}}
        }
        {
            \cE_{#1}\paren{{#2, #3}}
        }
    }
  }
}

\NewDocumentCommand\groupSeq{ m g }{
  \ensuremath{
    \IfNoValueTF{#2}
    {{\mathcal{S}}_{#1}}
    {{\mathcal{S}}_{#1, #2}}
  }
}
\NewDocumentCommand\unifiedGroupSeq{ m g }{
  \ensuremath{
    \IfNoValueTF{#2}
    {\bar{\mathcal{S}}_{#1}}
    {\bar{\mathcal{S}}_{#1, #2}}
  }
}
\NewDocumentCommand\DiscreteLapNoise{ m g }{
  \ensuremath{
    \IfNoValueTF{#2}
    {\operatorname{\mathbb{DL}ap}\left( {#1} \right)}
    {\operatorname{\mathbb{DL}ap}\left( {#1}, {#2} \right)}
  }
}
\NewDocumentCommand\TDiscreteLapNoise{ m g }{
  \ensuremath{
    \IfNoValueTF{#2}
    {\operatorname{\mathbb{TDL}ap}\left( {#1} \right)}
    {\operatorname{\mathbb{TDL}ap}\left( {#1}, {#2} \right)}
  }
}
\NewDocumentCommand\ExpNoise{ m g }{
  \ensuremath{
    \IfNoValueTF{#2}
    {\operatorname{\mathbb{E}xp}\left( {#1} \right)}
    {\operatorname{\mathbb{E}xp}\left( {#1}, {#2} \right)}
  }
}

\newif\ifcomment
\commenttrue

\definecolor{DarkGreen}{rgb}{0.1,0.5,0.1}
\ifcomment
\newcommand{\hao}[1]{\textcolor{blue}{[HAO: #1]}}

\else

\makeatletter
\newcommand{\hao}[1]{%
  \@bsphack
  \@esphack
}
\makeatother

\fi

\begin{document}

\title{Optimal Bounds for Private Minimum Spanning Trees via Input Perturbation}

\author{Rasmus Pagh}
\authornote{All authors contributed equally to this research.}
\orcid{xxx}
\affiliation{%
  \institution{BARC, University of Copenhagen}
  \country{Denmark}
}
\email{pagh@di.ku.dk}

\author{Lukas Retschmeier}
\authornotemark[1]
\affiliation{%
  \institution{BARC, University of Copenhagen}
  \country{Denmark}
}
\email{lure@di.ku.dk}

\author{Hao Wu}
\authornotemark[1]
\authornote{This work was partially carried out while the author was still at \emph{BARC, University of Copenhagen}.}
\affiliation{%
  \institution{University of Waterloo}
  \country{Canada}
}
    \email{hao.wu1@uwaterloo.ca}

\author{Hanwen Zhang}
\authornotemark[1]
\affiliation{%
  \institution{BARC, University of Copenhagen}
  \country{Denmark}
}
\email{hazh@di.ku.dk}

\renewcommand{\shortauthors}{Pagh et al.}

\begin{abstract}

We study the problem of privately releasing an approximate minimum spanning tree (MST).
Given a graph \( G = (V, E, \vec{W}) \) where \(V\) is a set of \( n \) vertices, \(E\) is a set of $m$ undirected edges, and \( \vec{W} \in \R^{|E|} \) is an edge-weight vector, our goal is to publish an approximate MST under \emph{edge-weight} differential privacy, as introduced by Sealfon in PODS 2016, where \(V\) and \(E\) are considered public and the weight vector is private.
Our neighboring relation is \(\ell_\infty\)-distance on weights: for a sensitivity parameter \(\Delta_\infty\), graphs \( G = (V, E, \vec{W}) \) and \( G' = (V, E, \vec{W}') \) are neighboring if \(\|\vec{W}-\vec{W}'\|_\infty \leq \Delta_\infty\).

Existing private MST algorithms face a trade-off, sacrificing either computational efficiency or accuracy.
We show that it is possible to get the best of both worlds: With a suitable random perturbation of the input that does \emph{not} suffice to make the weight vector private, the result of \emph{any} non-private MST algorithm will be private and achieves a state-of-the-art error guarantee.

Furthermore, by establishing a connection to \emph{Private Top-k Selection} [Steinke and Ullman, FOCS '17], we give the first privacy-utility trade-off lower bound for MST under approximate differential privacy, demonstrating that the error magnitude, \(\tilde{O}(n^{3/2})\), is optimal up to logarithmic factors.
That is, our approach matches the time complexity of any non-private MST algorithm and at the same time achieves optimal error.
We complement our theoretical treatment with experiments that confirm the practicality of our approach.
\end{abstract}


\received{Dec 2025}

\maketitle

\begin{table*}[t]
\centering
\resizebox{0.90\columnwidth}{!}{%
    \begin{tabular}{l|c|c|l}
    {\bf \newline Reference} & {\bf Additive Error} & {\bf Time}  & {\bf Technique}\\
    \hline
    \cite{Sealfon_2016} & $O \PAREN{ \paren{1 / \eps} \cdot n \sqrt{m} \cdot \paren{ \log n } \sqrt{\log \paren{1 / \delta}} }$ & MST + $O(m)$ & Input privatization\\
    \cite{Sealfon_2016} & $\Omega(n)$ & -- & Lower bound \\
    \cite{pinot_2018_ma} & $O \PAREN{  \paren{1 / \eps} \cdot n^{3/2} \cdot \paren{ \log n } \, \sqrt{\log \paren{1 / \delta}} }$ & $O(nm)$ & In-place noise\\
    \cite{pagh2024fasterprivateminimumspanning} & $O \PAREN{  \paren{1 / \eps} \cdot n^{3/2} \cdot \paren{ \log n } \, \sqrt{\log \paren{1 / \delta}} }$ & $O \left(m + n^{3/2} \, \log n \right)$ & In-place noise \\
    {\bf New} & $O \PAREN{  \paren{1 / \eps} \cdot n^{3/2} \cdot \paren{ \log n } \, \sqrt{\log \paren{1 / \delta}} }$ & MST + $O(m)$ & Input perturbation\\
    {\bf New} & $\Omega \PAREN{  \paren{1 / \eps} \cdot n^{3/2} \cdot { (\log n) } }$ & -- & Lower bound \\
    \hline
    \end{tabular}
}\\
\caption{
    Results on $(\eps, \delta)$-DP MST with $\ell_\infty$ neighboring relationship on edge weights, assuming sensitivity $\Delta_\infty = 1$. 
    \citet{Sealfon_2016} and \citet{pinot_2018_ma} originally provided upper bounds for pure DP, but these can be adapted for approximate DP. 
    \citet{Sealfon_2016} considered the $\ell_1$ neighboring relationship, which leads to the same lower bound and a weaker upper bound for $\ell_\infty$. 
    The unpublished manuscript \citep{pagh2024fasterprivateminimumspanning} was written by a subset of the authors. 
    The "MST" in the {\bf Time} column refers to the running time of any non-private MST algorithm.
}
\label{tab:summ}
\vspace{-7mm}
\end{table*}

\section{Introduction}

Graph algorithms are a cornerstone of data analysis, but directly applying classical graph algorithms to inputs that contain sensitive information risks disclosing too much information about the input.
In recent years \emph{differentially private} graph algorithms have emerged as a principled approach to ensuring that such disclosure is limited.
We refer to the survey of~\citet{li2023survey} for a general overview of private graph algorithms and to the tutorial of~\citet*{brito2024differentially} for a data management specific overview.

The \emph{minimum spanning tree} (MST) problem is a classical graph optimization problem with many applications.
With differential privacy constraints, it has been studied in the context of clustering algorithms~\cite{DBLP:journals/ida/LaiRN09,DBLP:conf/nips/BateniBDHKLM17,Pinot_2018,jayaram2024massively} and as a subroutine for computing graphical models~\cite{McKenna_Miklau_Sheldon_2021}.

We consider the problem of privately releasing an MST for a graph $G$ 
where both the set of vertices $V = \{1,\dots,n\}$ and edges $E = \set{1, \ldots, m}$ are public, but we keep the weight vector $\vec{W} = \paren{w_1, \ldots, w_m} \in \R^{|E|}$ private.
We consider the $\ell_\infty$ neighborhood relation, where all weights can differ by at most $\Delta_\infty$ and want to protect the information encoded on the weights, which is known as \emph{edge-weight differential privacy} \cite{Sealfon_2016}.

An application example is to model passenger data on a public city traffic network.
Another setting comes from graphical modeling: Consider a dataset $D = (\mathbf{x}^{(1)}, \ldots, \mathbf{x}^{(d)})$ of size $d$ where each vector $\mathbf{x}^{(i)}\in\{0,1\}^n$ represents a list of sensitive binary attributes. 
The \emph{Chow-Liu Tree}~\cite{chowliu1968} is the minimum spanning tree derived from the negated \emph{mutual information matrix} encoding all pairwise mutual information of the attributes.
Changing one $\vec{x}$ in $D$ could simultaneously alter \textit{all} weights.

Currently, there are two different algorithmic approaches: 
\emph{input-privatization} \cite{Sealfon_2016} and \emph{in-place} \cite{Pinot_2018,McKenna_Miklau_Sheldon_2021}.
The former method adds noise (e.g., Laplace or Gaussian) to each edge weight and publishes the entire noisy weight vector.
This allows anyone to compute the MST while privacy is ensured by the \emph{post-processing} property of differential privacy.
While this approach allows the design of an (expected) linear-time algorithm, it provides only a worst-case additive error of $\Tilde{O}(n^2)$.

The latter approach, which adds noise during the execution of an MST algorithm, achieves a better error bound of $\Tilde{O}(n^{3/2})$, but its best-known running time is $\Tilde{O}(n^{3/2} + m)$ for a fixed privacy parameter \cite{pagh2024fasterprivateminimumspanning}.
Thus, there is a gap, and whether it is possible to achieve the ``best of both worlds'' has remained unknown.
We answer these two open questions in the affirmative:

\begin{quote}
    \vspace{1mm}
    {\bf Question 1:} Can we design an (expected) linear-time private MST algorithm that matches the error guarantee of $\Tilde{O}(n^{3/2})$ of the \emph{in-place} approach?

    \vspace{2mm}
    {\bf Question 2:} Is the error asymptotically optimal?
\end{quote}

\subsection{Our Contribution}

We introduce the \emph{first} algorithmic framework that reduces the private MST problem (under the $\ell_\infty$ neighboring relationship) to the non-private one, achieving an error of $\Tilde{O}(n^{3/2})$. 
Compatible with any non-private MST algorithm, it enables the \emph{first} (expected) linear-time private MST solution. 
We also establish the \emph{first} $\Tilde{\Omega}(n^{3/2})$ error lower bound, proving the asymptotic optimality of our framework. Formally, we show the following upper and lower bounds:

\begin{restatable}[Upper Bound]{theorem}{upperbound}
\label{th:upper-bound}
    Let $G = (V, E, \vec{W})$ be a graph with $n$ vertices and $m$ edges, and let $\eps, \delta > 0$.
    Consider an arbitrary (non-private) algorithm that computes an MST of $G$ within time $t(n,m)$, independent of the weight vector.
    Then there exists an $\paren{\eps, \delta}$-differentially private mechanism $\cM$ that releases a random, approximately minimum spanning tree $\cM(G)$, such that, if $T^*$ is a minimum spanning tree:
    \begin{itemize}[leftmargin=5mm]
        \item the weight difference satisfies $\E{ \sum_{e \in \cM(G)} w_e } - \sum_{e\in T^*} w_e \in O \paren{ \frac{1}{\eps} \cdot n^{3/2} \cdot \paren{ \log n } \cdot \sqrt{\log \paren{1 / \delta}} }$, and
        \item the running time is $t(n,m) + O (m)$.
    \end{itemize}
\end{restatable}

For example, the randomized linear time algorithm \cite{karger1996} or the best-known near-linear time deterministic algorithm \cite{chazelle2000} can be used in \cref{th:upper-bound}. 
Our approach matches all known upper bounds, 
meanwhile it is very simple to implement. 
See~\cref{tab:summ} for a comprehensive summary.
Moreover, the result can be easily extended to finding a maximum weight independent set in any matroid with rank $n$ and $m$ elements, where $t(n, m)$ is the running time for any algorithm that computes the maximum weight independent set in such a matroid.

The key technique is a novel approach (formally introduced in Section~\ref{sec:one-pass-priv-kruskal}) to iteratively sample $k$ items from a larger candidate set with probabilities proportional to their weights. After each selection, the chosen item and a subset of the remaining candidates—determined by the sequence of previously selected items—are removed before the next selection.  
Our approach generates noisy weights for all items at the beginning of the sampling process, which are reused throughout the iterative selection procedure, thereby eliminating the need to regenerate fresh randomness at each step.  
As discussed in Section~\ref{sec:summary}, this technique extends a substantial body of research on top-$k$ sampling~\citep{YELLOTT1977, ohlsson1990sequential, ROSEN1997135, Cohen97}.

\vspace{1.5mm}
\noindent \textbf{Lower Bound. \,}
Our lower bound partially resolves an open question of \cite{hladik_tetek_2024}, establishing a tight bound for worst-case instances under approximate differential privacy:

\begin{theorem}[Lower Bound]
    \label{thm:lower bound of private MST}
    Let $\eps \leq 1$ and $\delta \in O \PAREN{1  /\sqrt{n}}$, and let $\cM$ be an $(\eps, \delta)$-DP MST algorithm.  
    There exists an input graph $G = (V, E, \vec{W})$ with $n$ vertices such that, if $T^*$ is the minimum spanning tree of $G$, the (random) spanning tree $\cM(G)$ released by $\cM$ satisfies:  
    \begin{align}
        \begin{array}{c}
            \E{\cM}{ \sum_{e \in \cM(G)} w_e } - \sum_{e \in T^*} w_e 
            \in \Omega \PAREN{ 
                \frac{1}{\eps} \cdot n^{3 / 2} \cdot \ln n
            }.
        \end{array}
    \end{align}
\end{theorem}

The proof technique is based on a lower bound technique for top-$k$ selection (reporting $k$ elements of approximately maximum value) in \citep{SteinkeU17}.
First, observe that maximization and minimization are equivalent for spanning trees by simply flipping the sign of all edge weights.
A maximum spanning tree can be seen as a top-$(n-1)$ selection problem (selecting $n-1$ edges of the approximately largest value) with an additional constraint that the edge set is acyclic.
Thus, it is natural to guess that the hard instances of \citep{SteinkeU17}, embedded into the weights of a complete graph, yield a hard instance for the maximum spanning tree as well.
By carefully adapting the proof in \citep{SteinkeU17}, we show that this is indeed the case.
In particular, leveraging a result from the Erdős–Rényi random graph model~\citep{erdos59a}, we show the existence of a spanning tree with weight similar to the top-$(n-1)$ edge weights.

\vspace{3mm}
\noindent 
{\bf Organization.} 
The rest of the paper is organized as follows. 
Section~\ref{sec:background} introduces the problem formally and reviews the necessary preliminaries.
Section~\ref{sec: upper bound} describes the reduction from normal non-private MST algorithm to private MST algorithm that achieve the desired utility and running time. 
Section~\ref{sec:lowerbound} establishes the lower bound for the problem. 
Section~\ref{sec:summary} presents a detailed comparison of our techniques with previous ones and explores the broader background of the problem.
Experiments in \cref{sec:empiricalevaluation} show the practicality of our approach.

\section{Preliminaries}\label{sec:background}

We consider a graph $G = (V, E, \vec{W})$ where the set of vertices $V = \{1,\dots,n\}$ and edges $E = \set{1, \ldots, m}$ are public, and the weight vector $\vec{W} = \paren{w_1, \ldots, w_m} \in \R^m$ is private.
For each edge $e \in E$, let $w_e$ denote its weight. 
The cost of a subset $T \subseteq E$ is defined as $w(T) := \sum_{e \in T} w_e$.
A {\textit{spanning tree}} is an acyclic subset $T \subseteq E$ that makes the graph connected.
Denote $\SpanningTrees{G}$ the collection of all possible spanning trees in $G$.
A \textit{minimum (cost) spanning tree} (MST) is a spanning tree $T^*$ which minimizes $w(T^*)$.
We sometimes hide logarithmic factors using tilde notation for $\tilde{O}$, $\tilde{\Theta}$ and $\tilde{\Omega}$.

\subsection{Differential Privacy} 

Given an input graph $G = (V, E, \vec{W})$, the private MST problem aims to find a spanning tree $T$ while preserving the privacy of the weight vector $\vec{W}$.  

\vspace{4mm}
\noindent {\bf Privacy Guarantee.}  
To achieve this, a private MST algorithm must produce similar output distributions for input graphs with similar weights.  
In what follows, we first formally define similar inputs and then similar output distributions.

\begin{definition}[$\ell_\infty$-neighboring inputs]
    Given $\Delta_\infty > 0$, two weight vectors $\vec{W}, \vec{W'} \subseteq \R^{E}$ are neighboring, 
    denoted $\vec{W} \sim \vec{W}'$, if and only if 
    $
        \norm{\vec{W} - \vec{W'}}_\infty = \max_{e \in E} \card{w_e - w_e'} \le \Delta_\infty.
    $
    Two graphs $G = (V, E, \vec{W})$ and $G' = (V', E', \vec{W}')$ are neighboring, denoted $G \sim G'$, 
    if and only if $V = V'$, $E = E'$, and $\vec{W} \sim \vec{W}'$.
\end{definition}

For simplicity, we assume throughout the paper that $\Delta_\infty = 1$. 
If $\Delta_\infty \neq 1$, all bounds presented can be generalized by scaling a factor of $\Delta_\infty$.

\begin{definition}[\citep{Dwork_Nissim_Smith_2006} $\paren{\eps, \delta}$-Private Algorithm] \label{def: Differential Privacy}
    Let $\eps, \delta > 0$, $G = (V, E)$ be a graph, and let $\SpanningTrees{G}$ denote the set of all possible spanning trees in $G$.
    An MST algorithm $\cM$
    is called~$\paren{\eps, \delta}$-differentially private (DP),
    if for every~$G = (V, E, \vec{W}), G' = (V', E', \vec{W}')$ such that~$G \sim G'$, 
    and all $Z \subseteq \SpanningTrees{G}$,
    \vspace{-2mm}
    \begin{equation} \label{ineq: def private algo}
        \begin{array}{c}
            \Pr[ \cM (G) \in Z ] \le e^\eps \cdot \Pr [ \cM (G') \in Z ] + \delta\,.
        \end{array}    
    \end{equation}
\end{definition}

The above definitions follow the notion of edge-weight differential privacy introduced by~\citet{Sealfon_2016}.  
Other privacy notions for graphs include \emph{edge-level privacy}~\cite{hay_li_miklau_jensen_2009} and \emph{node-level privacy}~\cite{Kasiviswanathan_nissim_sofya_smith_2013}.  

Further, although we present the definition in the context of private MST algorithms, it applies more generally to any randomized algorithms $\cM: \cX \rightarrow \cY$, where $\cX$ is the input space, which is associated with a symmetric relation $\sim$ that defines neighboring inputs.

\vspace{2mm}
\noindent {\bf Composition.}  
We also explore an alternative formulation of differentially privacy, which can offer a tighter analysis of the overall privacy guarantee of the composition of a sequence of private algorithms.

\begin{definition}[\citep{bun_steinke_2016} $\rho$-zero-Concentrated Differential Privacy]
Let $\rho > 0$.
An MST algorithm $\cM : \cX \rightarrow \cY$ satisfies $\rho$-zCDP, 
if for all $\alpha > 1$ and all pairs of neighboring inputs $X, X' \in \cX$, s.t., $X \sim X'$, it holds that 
\begin{equation}
    \begin{array}{c}
        D_\alpha \PAREN{ 
            \cM \PAREN{ X } || \cM \PAREN{ X' } 
        } 
        \le \rho \alpha, 
    \end{array}
\end{equation}
where 
$
    D_\alpha \PAREN{ 
        \cM \PAREN{ X } || \cM \PAREN{ X' } 
    }
$ 
denotes the $\alpha$-Rényi divergence between two output distributions of $\cM \PAREN{ X }$ and $\cM \PAREN{ X' }$.
\end{definition}

Its (partial) relationship with $(\eps, \delta)$-DP and its composition property are outlined below.  

\begin{fact}[\citep{bun_steinke_2016} Conversion]\label{fact:conversion-dp}  
    If $\cM$ satisfies $\rho$-zCDP, then $\cM$ is $(\rho + 2\sqrt{\rho \log(1/\delta)}, \delta)$-DP for all $\delta > 0$.  
    Conversely, if $\cM$ satisfies $\epsilon$-DP, then $\cM$ satisfies $\rho$-zCDP for $\rho \leq \frac{1}{2}\epsilon^2$.  
\end{fact}

\begin{fact}[\citep{bun_steinke_2016} Composition]\label{fact:composition-zdp}
    If $M_1$ and $M_2$ satisfy $\rho_1$-zCDP and $\rho_2$-zCDP, respectively,
    then $\cM=(\cM_1, \cM_2)$ satisfies $(\rho_1 + \rho_2)$-zCDP.
\end{fact}

\vspace{2mm}
\noindent {\bf Group Privacy.}
Group privacy studies the similarity between the output distributions of a private algorithm for two inputs that are "$r$-hops away", where $r \in \N^+$.

\begin{fact}[\citep{Vadhan17} Group Privacy]
    \label{fact: group privacy}
    Let $\cM : \cX \rightarrow \cY$ be an $(\eps, \delta)$-differentially private mechanism.  
    Given $r \in \N^+$ and $X, X' \in \cX$, if there exist $X^{(1)}, \ldots, X^{(r - 1)}$ such that $X^{(i - 1)} \sim X^{(i)}$ for each $i \in [r]$ (where we define $X^{(0)} = X$ and $X^{(r)} = X'$), then for each (measurable) subset $S \subseteq \cY$, we have:
    \vspace{-2mm}
    \begin{align}
        \begin{array}{c}
            \P{\cM(X) \in S} 
                \le e^{r \eps} \cdot \P{\cM(X') \in S}
                    + \frac{e^{r \eps} - 1}{e^{\eps} - 1} \cdot \delta.
        \end{array}
    \end{align}
\end{fact}

\vspace{2mm}

\noindent {\bf Exponential Mechanism.} 
The exponential mechanism~\citep{mcsherry2017} can be used to differentially private release discrete outputs.
It operates over an input space $\cX$ (with neighboring datasets defined by a relation $\sim$) and a finite output space $\cY$. 
The mechanism $\ExpoMesm: \cX \rightarrow \cY$ assigns the following probabilities for a given dataset $x \in \cX$:
\begin{align}
    \begin{array}{c}
 \P{\ExpoMesm(x) = y} 
            \propto \exp \bigparen{ - \eps \cdot \
            \loss{x}{y} \, / \, \paren{ 2 \cdot \SensitivityExpoMesm } },
        \quad \forall y \in \cY
    \end{array}
\end{align}
where $\loss: \cX \times \cY \rightarrow \R$ is the \emph{loss function} quantifying the cost of selecting $y$ given input $x$, and 
$\SensitivityExpoMesm$ is the \emph{sensitivity} of $\loss$, i.e., the maximum change in the loss function for neighboring datasets:
\[
    \SensitivityExpoMesm \doteq \max_{x \sim x', y \in \cY} \card{
            \loss{x}{y} - \loss{x'}{y}
        }.
\]

\begin{fact}[$\NameExpoMesm$~\citep{mcsherry2017} Properties of $\ExpoMesm$]  
    \label{fact: properties of exponential mechanism}  
    The exponential mechanism $\ExpoMesm$ satisfies $\eps$-differential privacy.  
    Moreover, for each $\beta \in \paren{0, 1}$, and
    $
        \tau \doteq \frac{2 \cdot \SensitivityExpoMesm}{\eps} \cdot \ln \frac{\card{\cY}}{\beta}, 
    $
    it holds that:
    \begin{equation}
        \begin{array}{cc} 
            \P{
                \loss{x}{\ExpoMesm(x)} \ge { \min_{y \in \cY} \, \loss{x}{y} } + \tau 
            } \le \beta,
            &
            \forall x \in \cX. 
        \end{array}
    \end{equation}
\end{fact}

\subsection{Probabilities}

We discuss several probability distributions applied in this paper.

\begin{definition}[\citep{Ross2018} Beta Distribution]
    The beta distribution~$\BetaDist{\alpha}{\beta}$ is a distribution defined on~$[0, 1]$ whose density is given by 
    $
        p(x) = \frac{
                x^{\alpha - 1} \paren{1 - x}^{\beta - 1}
            }{
                \BetaFunt{\alpha}{\beta}
            }, \, \forall x \in [0, 1], 
    $
    where~$\alpha, \beta > 0$ are \emph{shape parameters},~$\BetaFunt{\alpha}{\beta} \doteq \int_0^1 x^{\alpha - 1} \paren{1 - x}^{\beta - 1} \, dx$ is a normalization constant. 
\end{definition}

\noindent
The cornerstone of our algorithm design is the exponential distribution.

\begin{definition}[Exponential Distribution]
    \label{def:exponential distribution}
    Given $\lambda > 0$, a random variable $X$ follows exponential distribution $\ExpNoise{\lambda}$, denoted by $X \sim \ExpNoise{\lambda}$, if it has density
    $
        p_X(x) \doteq 
            \lambda e^{-\lambda x},\, \forall x \ge 0, 
    $
    and 
    $
        p_X(x) \doteq 0, \, \forall x < 0.
    $
\end{definition}

\vspace{2mm}
\section{Reduction from Private MST to Non-Private MST}
\label{sec: upper bound}

In this section, we present a novel algorithmic framework (Algorithm~\ref{algo:impl-details}) that enables a reduction from the private MST problem to the non-private one, thereby proving Theorem~\ref{th:upper-bound}.  
A key feature of this framework is its independence from the chosen MST algorithm. 
Since the graph's topology is public, the most suitable algorithm can always be selected in advance.  
Notably, this framework facilitates an (expected) linear-time private MST algorithm.  

\vspace{2mm}
\noindent {\bf Overview.}  
Figure~\ref{fig:roadmap} outlines the roadmap of this section.  
In Section~\ref{sec:algorithmicFramework}, we introduce a simple algorithmic framework that that fulfills Theorem~\ref{th:upper-bound}.
The framework adds certain noise to each edge weight, then applies an arbitrary non-private MST algorithm to publish only the set of MST edges with respect to the noisy weights. 
Proving its time complexity and utility is straightforward, but its privacy guarantee requires more effort.  
Instead of proving privacy directly, we establish it through two equivalent algorithms.  
Section~\ref{sec:privateKruskal} describes a private version of Kruskal's algorithm~\cite{kruskal_1956} (Algorithm~\ref{alg:privKruskal}), which iteratively and privately selects the approximate MST edges. 
Its privacy analysis follows directly from the composition property of DP, though its utility analysis is more intricate.  
Finally, in Section~\ref{sec:one-pass-priv-kruskal}, we present a "bridging" algorithm (Algorithm~\ref{alg:one-pass-private-kruskal}) that has the same output distribution as both Algorithm~\ref{algo:impl-details} and Algorithm~\ref{alg:privKruskal}, establishing their equivalence. 
This allows us to transfer privacy and utility guarantees between the algorithms, simplifying the proofs.

A detailed comparison of our algorithmic techniques with prior work, along with a broader discussion of the problem's background, is deferred to Section~\ref{sec:summary}.

\subsection{Reduction}
\label{sec:algorithmicFramework}

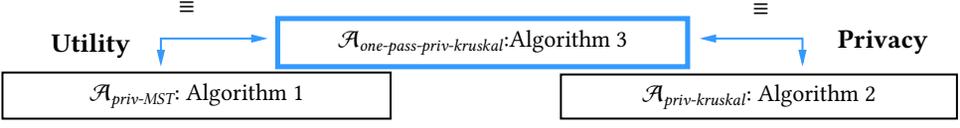
\begin{figure}
    \centering
    \tikzset{every picture/.style={line width=0.75pt}} %

\begin{tikzpicture}[x=0.75pt,y=0.75pt,yscale=-1,xscale=1]

\draw  [color={rgb, 255:red, 51; green, 153; blue, 255 } ]  (220,126) -- (220,114) -- (272,114) ;
\draw [shift={(274,114)}, rotate = 180] [fill={rgb, 255:red, 51; green, 153; blue, 255 }  ][line width=0.08]  [draw opacity=0] (8.4,-2.1) -- (0,0) -- (8.4,2.1) -- cycle    ;
\draw [shift={(220,128)}, rotate = 270] [fill={rgb, 255:red, 51; green, 153; blue, 255 }  ][line width=0.08]  [draw opacity=0] (8.4,-2.1) -- (0,0) -- (8.4,2.1) -- cycle    ;
\draw [color={rgb, 255:red, 51; green, 153; blue, 255 } ] (544,126) -- (544,114) -- (494,114) ;
\draw [shift={(492,114)}, rotate = 360] [fill={rgb, 255:red, 51; green, 153; blue, 255 }  ][line width=0.08]  [draw opacity=0] (8.4,-2.1) -- (0,0) -- (8.4,2.1) -- cycle    ;
\draw [shift={(544,128)}, rotate = 270] [fill={rgb, 255:red, 51; green, 153; blue, 255 }  ][line width=0.08]  [draw opacity=0] (8.4,-2.1) -- (0,0) -- (8.4,2.1) -- cycle    ;

\draw  [color={rgb, 255:red, 51; green, 153; blue, 255 }  ,draw opacity=1 ][line width=1.9]   (279,104.04) -- (486,104.04) -- (486,128.04) -- (279,128.04) -- cycle  ;
\draw (382.5,116.04) node  [font=\small] [align=left] {\begin{minipage}[lt]{138.11pt}\setlength\topsep{0pt}
\begin{center}
$\onePassPrivateKruskal$:\cref{alg:one-pass-private-kruskal}
\end{center}

\end{minipage}};
\draw    (140,130.96) -- (336,130.96) -- (336,153.96) -- (140,153.96) -- cycle  ;
\draw (238,142.46) node  [font=\small] [align=left] {\begin{minipage}[lt]{130.75pt}\setlength\topsep{0pt}
\begin{center}
$\privateMSTFramework$: \cref{algo:impl-details}
\end{center}

\end{minipage}};
\draw    (422,131.93) -- (622,131.93) -- (622,153.93) -- (422,153.93) -- cycle  ;
\draw (522,142.93) node  [font=\small] [align=left] {\begin{minipage}[lt]{133.47pt}\setlength\topsep{0pt}
\begin{center}
$\privateKruskal$: \cref{alg:privKruskal}
\end{center}

\end{minipage}};
\draw (560,108) node [anchor=north west][inner sep=1pt]   [align=left] {\textbf{Privacy}};
\draw (163,109) node [anchor=north west][inner sep=1pt]   [align=left] {\textbf{Utility}};
\draw (227,93.4) node [anchor=north west][inner sep=1pt]    {$\equiv$};
\draw (517,94.4) node [anchor=north west][inner sep=1pt]    {$\equiv$};

\end{tikzpicture}
    \caption{
        Roadmap of Section~\ref{sec: upper bound}. 
        The figure outlines the proof structure: the \emph{utility guarantee} is established for Algorithm~\ref{algo:impl-details}, and the \emph{privacy guarantee} is proven for Algorithm~\ref{alg:privKruskal}. 
        A "bridging" algorithm (Algorithm~\ref{alg:one-pass-private-kruskal}) is introduced to demonstrate the equivalence of Algorithms~\ref{algo:impl-details} and~\ref{alg:privKruskal} by showing they share the same output distribution.
    }
    \label{fig:roadmap}
\end{figure}

The framework is outlined in Algorithm~\ref{algo:impl-details}. 
It is straightforward to implement and requires only $3$ lines of code using standard libraries. 
Given a privacy parameter $\rho$ and an input graph $G = (V, E, \vec{W})$, it computes for each edge $e \in E$ a noisy weight $\tilde{w}_e \doteq w_e + \PAREN{2 / \eps'} \cdot \ln \ExpNoise{1}$, where \(\ExpNoise{1}\) denotes a random variable following the exponential distribution (\cref{def:exponential distribution}). 
Finally, it computes the MST using an existing (non-private) MST algorithm~$\cA_{MST}$
on the noisy weights $\tilde{\vec{W}} = \PAREN{\tilde{w}_1, \ldots, \tilde{w}_m}$.

\begin{algorithm}[H]
    \caption{\privateMSTFramework: Private MST Framework }
    \label{algo:impl-details}
    \begin{algorithmic}[1]
        \Statex \hspace{-4.5mm} {\bf Input: } $G = (V, E, \vec{W})$, any MST algorithm $\cA_{MST}$, privacy parameters $\PAREN{\eps, \delta}$
        \State $\rho \leftarrow \bigparen{
            \sqrt{\eps + \log \paren{1 / \delta}} - \sqrt{\log \paren{1 / \delta} \, }
        }^2$;\, 
        $
            \eps' \leftarrow \sqrt{{2 \cdot \rho} / \, \paren{n - 1}}
        $;\, 
        \label{line:algo:impl-details:initialization}
        \State $\tilde{w}_e \leftarrow w_e + \PAREN{2 / \, \eps'} \cdot \ln \left(\ExpNoise{1}\right)$ for all $e\in E$
        \label{line: algo:impl-details: noisy edge weights}
        \State \textbf{return} $\cA_{MST}(V,E, \tilde{\vec{W}})$
    \end{algorithmic}
\end{algorithm}

\noindent
{\it Extension to finding a maximum weight independent set in matroids:}  
The framework extends to general matroids by perturbing each element's weight with similar noise and computing the maximum weight independent set using any suitable algorithm on the noisy weights, see \cref{appendix:matroid}.

\vspace{2mm}
\noindent
The properties of \cref{algo:impl-details} are outlined in \cref{th:upper-bound}. 

\vspace{2mm}
\noindent {\it Proof of \cref{th:upper-bound}.} \,
It remains to analyze its running time, privacy and utility guarantees.

\noindent \textbf{Running Time.}
Adding noise to all edge weights requires only $O(m)$ time. 
Consequently, the overall running time of \cref{algo:impl-details} is given by $O(m + t_{\cA_{MST}}(n, m))$, where $t_{\cA_{MST}}(n, m)$ denotes the running time of the chosen MST algorithm $\cA_{MST}$ for a graph with $n$ vertices and $m$ edges. 
For example, with the celebrated MST algorithm by~\citet{karger1996}, \cref{algo:impl-details} achieves an expected linear running time, where $t_{\cA_{MST}}(n, m) \in O(n + m)$. 
Our reduction is compatible with any MST algorithm, even in parallel or distributed settings.

\vspace{3mm}
\noindent \textbf{Utility Guarantee.}
We analyze the utility of \cref{algo:impl-details} before establishing the privacy guarantee, as the former is more straightforward.
We need the following fact, whose proof is in Appendix~\ref{appendix: proofs for premilinary}.

\begin{fact}
   \label{fact: confidence interval of gumbel noise}
    Let $Z_i \sim \ln \ExpNoise{1}$ for $i \in [m]$ be independent.
    Then 
    $
        \E{ \max_{i \in [m]} |Z_i| } \in O \PAREN{ \ln m }.
    $
\end{fact}

\noindent
Let $T^*$ and $T$ denote the MSTs with respect to the original edge weights $\vec{W}$ and the noisy edge weights $\tilde{\vec{W}}$, respectively. 
Note that $T$ is precisely the spanning tree returned by Algorithm~\ref{algo:impl-details}. 
We aim to prove that $T$ satisfies the utility guarantee specified in Theorem~\ref{th:upper-bound}.

By Fact~\ref{fact: confidence interval of gumbel noise} and noting that both $T$ and $T^*$ contain fewer than $n$ edges, it follows that 
\begin{equation}
    \begin{array}{cc}
        \E{ \max_{e \in E} \card{ \tilde{w}_e - w_e } } \in O \PAREN{ {1 / \, \eps'} \cdot \ln m }
        \implies
        \begin{array}{l}
            \E{ \card{ w(T) - \tilde{w}(T) } }      \\ 
            \E{ \card{ w(T^*) - \tilde{w}(T^*) } }
        \end{array}
            \in O \PAREN{ {n / \, \eps'} \cdot \ln m }.
    \end{array}
\end{equation}
As $T$ is the MST on the noisy weights $\tilde{\vec{W}}$, we have $\tilde{w}(T) \le \tilde{w}(T^*)$. Therefore,
\begin{equation}
    \label{ineq:semi-final-utility-bound-of-algo:impl-details}
    \begin{array}{rl}
         \E{ w(T) - w(T^*) } 
            &\hspace{-3mm} \le \E{ w(T) - \tilde{w}(T) + \tilde{w}(T^*) - w(T^*) } \\
            &\hspace{-3mm} \le \E{ \card{ w(T) - \tilde{w}(T) } } + \E{ \card{ \tilde{w}(T^*) - w(T^*) } }
            \in O \PAREN{ {n / \, \eps'} \cdot \ln m }.
    \end{array}
\end{equation}
Finally, to obtain the utility bound in Theorem~\ref{th:upper-bound}, based on Algorithm~\ref{algo:impl-details}, Line~\ref{line:algo:impl-details:initialization}, we observe that
\begin{align}
    \hspace{-1mm}
    \begin{array}{cc}
        \frac{1}{\eps'} 
            = \sqrt{ \frac{n - 1}{2\rho} } 
            \in O \Bigparen{
                \frac{1}{\eps} \sqrt{n\log \frac{1}{\delta} }
            },
        & \hspace{-2mm} \text{since} \hspace{1mm}
        \frac{1}{ \sqrt{\rho} }
            = \frac{1}{
                \sqrt{\eps + \log {\frac{1}{\delta}} } - \sqrt{\log {\frac{1}{\delta}} \, }
            }
            = \frac{
                \sqrt{\eps + \log {\frac{1}{\delta}}} + \sqrt{\log {\frac{1}{\delta}} \, }
            }{\eps}
            \in O \Bigparen{
                \frac{1}{\eps} \sqrt{\log \frac{1}{\delta} }
            }.
    \end{array}
    \hspace{-3mm}
\end{align} 
Substituting the expression for $1 / \eps'$ into \cref{ineq:semi-final-utility-bound-of-algo:impl-details} completes the proof.

\vspace{3mm}
\noindent \textbf{Privacy Guarantee.}
As outlined in the \emph{roadmap} and illustrated in Figure~\ref{fig:roadmap}, directly proving the privacy guarantees is challenging.  
To address this, we introduce Algorithm~\ref{alg:privKruskal} in Section~\ref{sec:privateKruskal}, whose privacy guarantee can be easily established, and Algorithm~\ref{alg:one-pass-private-kruskal} in Section~\ref{sec:one-pass-priv-kruskal}, which shares the same output distribution as Algorithms~\ref{algo:impl-details} and~\ref{alg:privKruskal},   
transfering the privacy guarantee to the former.

\vspace{-4mm}
\hfill $\square$

\subsection{Private Kruskal Algorithm}\label{sec:privateKruskal}

Algorithm~\ref{alg:privKruskal} presents a privatized version of Kruskal's algorithm. 
Compared to the non-private one \cite{kruskal_1956}, at each iteration (Line~\ref{line: private kruskal start}-\ref{line: private kruskal end}), instead of selecting the edge with  minimum weight, it selects an approximately minimum-weight edge by the exponential mechanism~\citep{mcsherry2017} which samples an edge $e$ with probability proportional to $e^{- (\eps' / 2) \cdot w_e}$.

\begin{algorithm}[!h]
    \caption{$\cA_{\text{private-Kruskal}}$: Private Kruskal's Algorithm}
    \label{alg:privKruskal}
    \begin{algorithmic}[1]
        \Require Graph $G = (V, E, \vec{W})$, privacy parameter $\PAREN{\eps, \delta}$
        \State $\rho \leftarrow \bigparen{
            \sqrt{\eps + \log \paren{1 / \delta}} - \sqrt{\log \paren{1 / \delta} \, }
        }^2$;\, 
        $
            \eps' \leftarrow \sqrt{{2 \cdot \rho} / \, \paren{n - 1}}
        $;\, 
        \label{line:alg:privKruskal:initialization}
        \State Initialize $T \leftarrow \varnothing$, $F \leftarrow E$
        \For{$i \leftarrow 1$ to $|V| - 1$}
            \label{line: private kruskal start}
            \State Sample $e\in F$ with probability $\propto e^{- \frac{\eps'}{2} \cdot w_e}$
            \label{line: edge sampling}
            \State $T \leftarrow T \cup \set{ e }$
            \State $F \leftarrow F \setminus \set{e' \in F : \{ e' \} \cup T \text{ contains a cycle} }$
            \label{line: private kruskal end}
        \EndFor
        \State \textbf{return} $T$
    \end{algorithmic}
\end{algorithm}

\begin{restatable}[Properties of \cref{alg:privKruskal}]{theorem}{utilityPrivateKruskal}\label{theorem:properties-of-PrivateKruskal}
Given an input graph $G = (V, E, \vec{W})$ with $n$ vertices and $m$ edges, 
Algorithm~\ref{alg:privKruskal} runs in $\mathcal{O}(m + n \log n)$ time, is $(\eps, \delta)$-DP, 
and returns an approximately minimum spanning tree $\cM(G)$ such that, if $T^*$ is a minimum spanning tree:
\begin{align}
    \begin{array}{c}
        \E{ \sum_{e \in \cM(G)} w_e } - \sum_{e\in T^*} w_e \in O \paren{ \frac{1}{\eps} \cdot n^{3/2} \cdot \paren{ \log n } \cdot \sqrt{\log \paren{1 / \delta}} }.
    \end{array}
\end{align}
            
\end{restatable}

\noindent
{\it Proof of Theorem~\ref{theorem:properties-of-PrivateKruskal}.} \,
We analyze the running time, privacy and utility guarantees of Algorithm~\ref{alg:privKruskal}.

\vspace{2mm}
\noindent
\textbf{Running Time.}
\cref{alg:privKruskal} can be implemented in $\mathcal{O}(m + n \log n)$ time using a designated data structure. 
As it is not our primary focus, we refer for the details to \cref{appx:implPrivKruskal}.
It remains to analyze the privacy-utility trade-offs of Algorithm~\ref{alg:privKruskal}.

\vspace{2mm}
\noindent
\textbf{Privacy Guarantee.}
We leverage zCDP for its tighter composition bounds, so the proof involves conversion from DP to zCDP and back.
We need to show that Algorithm~\ref{alg:privKruskal} guarantees $(\eps, \delta)$-DP.  
\begin{itemize}[leftmargin=*]
    \item Based on the privacy guarantee of the exponential mechanism (Fact~\ref{fact: properties of exponential mechanism}), each round of Algorithm~\ref{alg:privKruskal} is $\eps'$-DP.  
    Using the conversion from DP to zCDP (\cref{fact:conversion-dp}) and the initialization of $\eps'$ (Line~\ref{line:alg:privKruskal:initialization}), this corresponds to $\rho'$-zCDP, where $\rho' = {\PAREN{\eps'}^2 / 2} = \rho / (n - 1)$. 

    \item Since the algorithm runs $n - 1$ rounds, by the composition property (\cref{fact:composition-zdp}), Algorithm~\ref{alg:privKruskal} satisfies $(n - 1) \cdot \rho' = \rho$-zCDP.  

    \item Applying \cref{fact:conversion-dp} in reverse (from zCDP to DP), we conclude that the algorithm satisfies $(\rho + 2 \cdot \sqrt{\rho \log(1/\delta)}, \delta)$-DP.  
    Substituting $\rho$ with  
    $
        \paren{\sqrt{\eps + \log \paren{1 / \delta}} - \sqrt{\log \paren{1 / \delta}}}^2
    $
    yields  
    $
        \rho + 2 \cdot \sqrt{\rho \log(1/\delta)} = \eps.
    $
\end{itemize}

\vspace{2mm}
\noindent
\textbf{Utility Guarantee.}
The utility analysis is more involved, so we defer it to the next section, where we introduce a simplified implementation of Algorithm~\ref{alg:privKruskal} and establish its equivalence to Algorithm~\ref{algo:impl-details}.  
This allows the straightforward utility analysis of Algorithm~\ref{algo:impl-details} to transfer directly to Algorithm~\ref{alg:privKruskal}, demonstrating the benefit of proving the algorithms' equivalence.

It is worth noting that the utility guarantee of Algorithm~\ref{alg:privKruskal} can be proven directly, although it is much more complicated and requires a much deeper insight into Kruskal’s algorithm.  
A proof is provided in \cref{proof:utilityPrivateKruskal} for reference.

\hfill $\square$

\subsection{One-Pass Private Kruskal}\label{sec:one-pass-priv-kruskal}

Finally, we introduce the last private MST algorithm. 
It achieves the same running time as Algorithm~\ref{alg:privKruskal}, but features a significantly simplified structure, allowing it to be implemented using existing libraries for non-private Kruskal's algorithm. 
More importantly, it serves as a bridge to establish the equivalence between Algorithm~\ref{algo:impl-details} and Algorithm~\ref{alg:privKruskal}.

The algorithm is outlined in Algorithm~\ref{alg:one-pass-private-kruskal}. 
It begins by generating noisy edge weights using $\tilde{w}_e \doteq \ExpNoise{1} / e^{- (\eps' / 2) w_e}$ and then applies the non-private Kruskal's algorithm to the noisy weights.

\begin{algorithm}[!h]
    \caption{$\cA_{\text{one-pass-private-Kruskal}}$}
    \label{alg:one-pass-private-kruskal}
    \begin{algorithmic}[1]
        \Require Graph $G = (V, E, \vec{W})$, privacy parameter $\PAREN{\eps, \delta}$
        \State $\rho \leftarrow \bigparen{
            \sqrt{\eps + \log \paren{1 / \delta}} - \sqrt{\log \paren{1 / \delta} \, }
        }^2$;\, 
        $
            \eps' \leftarrow \sqrt{{2 \cdot \rho} / \, \paren{n - 1}}
        $;\, 
        \State Initialize $T \leftarrow \varnothing$, $F \leftarrow E$
        \State $\tilde{w}_e \leftarrow \ExpNoise{1} / e^{- \frac{\eps'}{2} \cdot w_e}$ \Comment{$\ExpNoise{1}$ represents exponential noise}
        \For{$i \leftarrow 1$ to $|V| - 1$}
            \State $e \leftarrow \arg\min_{e \in F} \tilde{w}_e$ 
            \label{line: pick edge in alg:one-pass-private-kruskal}
            \State $T \leftarrow T \cup \set{ e }$ \label{line:remove}
            \State $F \leftarrow F \setminus \set{e' \in F : \{ e' \} \cup T \text{ contains a cycle} }$
            \label{line: removing cycling edges in alg:one-pass-private-kruskal}
        \EndFor
        \State \textbf{return} $T$
    \end{algorithmic}
\end{algorithm}

\begin{theorem}%
    \label{theorem:property-one-pass-private-kruskal}
    \cref{alg:one-pass-private-kruskal} shares the properties of \cref{alg:privKruskal} as stated in \cref{theorem:properties-of-PrivateKruskal}.
\end{theorem}

\noindent
{\it Proof of Theorem~\ref{theorem:property-one-pass-private-kruskal}.} \,
We analyze the runtime, utility, and privacy guarantees of Algorithm~\ref{alg:one-pass-private-kruskal}.  
In the process, we establish the equivalence among \cref{algo:impl-details}, \cref{alg:privKruskal}, and \cref{alg:one-pass-private-kruskal}, i.e., they have the same output distributions.

\vspace{1mm}
\noindent \textbf{Running Time.}
The running time is dominated by Kruskal's algorithm, which is $O(n + m \log n)$.

\vspace{1mm}
\noindent \textbf{Utility Guarantee.}
The utility guarantee is established through the equivalence between \cref{algo:impl-details} and \cref{alg:one-pass-private-kruskal}, allowing the utility guarantee of the former to transfer to the latter.

First, observe that in line~\ref{line: pick edge in alg:one-pass-private-kruskal} of \cref{alg:one-pass-private-kruskal}, the returned edge satisfies
\begin{align}
    \begin{array}{c}
        \argmin_{e \in F} \tilde{w}_e 
            = \argmin_{e \in F} \PAREN{ w_e + \frac{2}{\eps'} \cdot \ln \ExpNoise{1} }.
    \end{array}
\end{align}
Hence, we can replace the noisy weight $\tilde{w}_e \doteq \ExpNoise{1} / e^{- \frac{\eps'}{2} \cdot w_e}$ in \cref{alg:one-pass-private-kruskal} by $\tilde{w}_e \doteq w_e + \frac{2}{\eps'} \cdot \ln \ExpNoise{1}$.
Second, since $\frac{2}{\eps'} \cdot \ln \ExpNoise{1}$ is a continuous random variable, the MST on the noisy graph is unique with probability $1$.
Consequently, \emph{any} MST algorithm computes the same solution, thereby establishing the equivalence between \cref{algo:impl-details} and \cref{alg:one-pass-private-kruskal}.

\vspace{1mm}
\noindent \textbf{Privacy Guarantee.}
The privacy guarantee is established through the equivalence between \cref{alg:privKruskal,alg:one-pass-private-kruskal}, allowing the privacy guarantee of the former to transfer to the latter.

To establish the equivalence, we introduce a generalized model, \emph{Probability Proportional to Sizes with Adaptive Candidate Removal} (\PPSACR), for top-$k$ sampling, along with a new sampling technique for this model. 
As will be shown, \cref{alg:privKruskal} can be viewed as a special case of \PPSACR, and \cref{alg:one-pass-private-kruskal} directly implements the new sampling technique, thereby demonstrating its equivalence to \cref{alg:privKruskal}.
It is worth noting that the equivalence between \cref{alg:privKruskal,alg:one-pass-private-kruskal} can also be established directly, without relying on \PPSACR. 
However, we choose to present \PPSACR here, as it is of independent interest and may have broader applications to other problems.

\subsubsection*{\PPSACR Model:}
\label{sec:iter-exp-mechanism}

The model is described in \cref{alg:iterative}.  
Let $U$ be a finite set of items, indexed from $1$ to $|U|$. Each item $j \in U$ is associated with a weight $s(j) \in \R^+$. \PPSACR iteratively samples $k$ items. It maintains a candidate set $\cC$, initialized as $U$, and a selected item list $\cI$, which is initially empty.  
At each step, an item $j \in \cC$ is sampled with probability proportional to $s(j)$ and appended to $\cI$. Afterward, the sampled item $j$ and an additional subset of items, denoted by $f(\cI)$, are removed from $\cC$. Here, $f : \PAREN{\bigcup_{i \in [k]} U^i} \rightarrow 2^{U}$ is a function that determines the additional items to be removed based on the sequence of selected items so far.  
Since the set $U \setminus \cC$ at each step is fully determined by $\cI$ and $f$, the function $f$, given $\cI$ as input, effectively encodes the information about $U \setminus \cC$.

\begin{example}
    \label{example:private kruskal as PPSACR}
    \cref{alg:privKruskal} can be interpreted as a special case of this sampling model, where $U$ represents the set of edges $E$, $s(e)$ is the exponential mechanism sampling weight $e^{-\lambda \cdot w_e^*}$ for each $e \in E$, $k$ is $n - 1$ (the number of edges in a spanning tree), and $f$ is the function that removes edges forming cycles based on the previously selected edges.
\end{example}

\begin{algorithm}[!h]
    \caption{\PPSACR}
    \label{alg:iterative}
    \begin{algorithmic}[1]
        \State $\cC \leftarrow U, \, \cI \leftarrow$ An Empty List \Comment{possible candidates and sampled items so far}
        \label{line: alg:iterative: initialization}
        \While{$|\cI| \le k$ and $\cC \neq \varnothing$}
            \State Sample $j \in \cC$ with probability proportional to $s(j)$
            \label{line: alg:iterative: sampling}
            \State Add $j$ to the back of $\cI$
            \State $\cC \gets \cC \setminus \PAREN{ \set{ j } \cup f \big( \cI \big) }$
            \label{line: alg:iterative: remove}
        \EndWhile
        \State \textbf{return} $\cI$
    \end{algorithmic}
\end{algorithm}

\begin{theorem}
    \label{theorem: ppsacr linera randomness}
    \cref{alg:iterative} can be equivalently implemented with the following modifications: 
    \begin{enumerate}
        \item Add a noisy weights generating step before line~\ref{line: alg:iterative: initialization}: $\Tilde{s}(j) \leftrightarrow \ExpNoise{1} / s(j), \forall j \in U.$

        \item Replace line~\ref{line: alg:iterative: sampling} with $j \leftarrow \argmin_{j' \in \cC} \Tilde{s}(j')$.
    \end{enumerate}
    \hspace{2.8mm} The modified algorithm is referred to as \OneShotPPSACR.  
\end{theorem}

Due to space limit, the pseudo codes of the modified algorithm (\cref{alg:one-shot-PPSACR}) is given in \cref{appendix: One-Shot-PPSACR}.
Note that if \cref{alg:privKruskal} is viewed as a special case of \PPSACR, as described in \cref{example:private kruskal as PPSACR}, then \cref{alg:one-pass-private-kruskal} directly implements the modifications proposed in \cref{theorem: ppsacr linera randomness}.  
\emph{
    Therefore, \cref{theorem: ppsacr linera randomness} immediately establishes the equivalence between \cref{alg:privKruskal,alg:one-pass-private-kruskal}.
}

To proceed, we require the following properties of the exponential distribution, the proofs of which are included in Appendix~\ref{appendix: proofs for premilinary}.

\begin{fact}[Scaling]\label{fact: scaled exponential}
     If $X \sim \ExpNoise{1}$, then for all $\lambda > 0$, then $Y \doteq X / \lambda$ has distribution $\ExpNoise{\lambda}$.
\end{fact}
\vspace{-4mm}
\begin{fact}[Minimum]\label{fact:min of exps}
    If $X_1 \sim \ExpNoise{\lambda_1}, \ldots, X_d \sim \ExpNoise{\lambda_d}$, then $\P{ X_i = \min_{j \in [d]} X_{j}} = {\lambda_i} / \paren{\sum_{j \in [d]} \lambda_j}$.
\end{fact}
\vspace{-4mm}
\begin{fact}[Memoryless]\label{fact:memoryless of exps}
    If $X \sim \ExpNoise{\lambda}$, then 
    $
        \P{X \ge x + y \mid X \ge x} = \P{ X \ge y },\,  \forall x, \, y \ge 0.
    $
\end{fact}

\begin{proof}[Proof of Theorem~\ref{theorem: ppsacr linera randomness}]
    To simplify the discussion, we assume that $k$ sampled items are returned by \PPSACR. 
    The proof for the case when less than $k$ items are returned are similar. 
    Let $\cJ \doteq (j_1, j_2, \ldots, j_k) \in U^k$ be a feasible output sequence of \PPSACR.
    We will prove that, \OneShotPPSACR outputs $\cJ$ with the same probability as \PPSACR.

    \vspace{2mm}
    \noindent {\it \PPSACR:} 
    For each $i \in [k]$, denote $\cC_{i}$ the set of candidates to be sampled in \cref{alg:iterative}, Line~\ref{line: alg:iterative: sampling} during the $i^{(th)}$ iteration.
    The probability of selecting $j_i$ is $s(j_i) / \bigparen{ \sum_{j \in \cC_i} s(j) }$. 
    Thus, the probability that \PPSACR outputs $\cJ$ is 
    \begin{equation} 
        \label{eq: probability of alg:iterative}
            \frac{s(j_1)}{ \sum_{j \in \cC_1} s(j)  }
            \cdot 
            \frac{s(j_2)}{ \sum_{j \in \cC_2} s(j)  }
            \cdots
            \frac{s(j_k)}{ \sum_{j \in \cC_k} s(j)  }.
    \end{equation}

    \vspace{2mm}
    \noindent {\it \OneShotPPSACR:} 
    For each $i \in [k]$, let $\cE_i$ be the event that $\tilde{s}(j_i) = \min_{j \in \cC_i} \, \tilde{s}(j)$, and define $\cE_{1 : i} \doteq \cE_1 \wedge \ldots \wedge \cE_i.$
    First, by \cref{fact: scaled exponential}, for each $j \in U$, $\tilde{s}(j) \doteq \ExpNoise{1} / s(j)$ follows distribution $\ExpNoise{s(j)}$.
    Based on \cref{fact:min of exps}, the minimum property of a collection of exponential random variables, it holds that 
    \begin{align}
        \P{ \cE_1 } = \frac{s(j_1)}{ \sum_{j \in \cC_1 } s(i) }.
    \end{align}
    
    Second, let $z_1 \le z_2 \le \ldots \le z_k$ be a feasible realization of $\tilde{s}(j_1), \ldots, \tilde{s} (j_k).$
    For each $i \in \IntSet{2}{k}$, 
    define $\tilde{s}(j_{1 : i - 1}) \doteq \PAREN{\tilde{s}(j_{1}), \ldots, \tilde{s}(j_{i - 1})}$ and $z_{1 : i - 1} = \PAREN{z_1, \ldots, z_{i - 1}}$.
    Conditioned on the events $\cE_{1 : i - 1}$ and $\tilde{s}(j_{1 : i - 1}) = z_{1 : i - 1}$, we know that for each $j \in \cC_i$, $\tilde{s}(j) \ge \PAREN{\max z_{1 : i - 1}} = z_{i - 1}$.  
    Since the $\tilde{s}(j)$ are independent, by the memoryless property of the exponential distribution (\cref{fact:memoryless of exps}),  
    $\tilde{s}(j) - z_{i - 1}$ still follows the distribution $\ExpNoise{s(j)}$.  
    As the event $\cE_i$ is now equivalent to $\tilde{s}(j_i) - z_{i - 1} = \min_{j \in \cC_i} \PAREN{\tilde{s}(j) - z_{i - 1}}$, applying \cref{fact:min of exps} again yields:  
    \vspace{-2mm}
    \begin{align}
        \P{ 
            \cE_i
            \, | \, 
            \cE_{1 : i - 1}, \, 
            \tilde{s}(j_{1 : i - 1}) = z_{1 : i - 1}
        } 
        = \frac{s(j_i)}{ \sum_{j \in \cC_i} s(j)  }.
    \end{align}
    Taking expectation over $\tilde{s}(j_{1 : i - 1})$ gives 
    $
        \E{
            \tilde{s}(j_{1 : i - 1})
        }{
            \P{ 
                \cE_i
                \, | \, 
                \cE_{1 : i - 1}, \, 
                \tilde{s}(j_{1 : i - 1})
            } 
        }
        = \P{ 
                \cE_i
                \, | \, 
                \cE_{1 : i - 1}
            } 
        = \frac{s(j_i)}{ \sum_{j \in \cC_i} s(j)  }.
    $
        
    Finally, based on chain rule of probability, $\cE_{1 : k}$ has exactly the same probability as \cref{eq: probability of alg:iterative}:
    
    \vspace{-4mm}
    \begin{align}
        \hspace{-5mm}
        \P{ 
            \cE_{1 : k}
        } 
        &= \P{ 
            \cE_{1}
        } 
        \cdot \prod_{i = 2}^k
        \P{ 
            \cE_i
            \, | \, 
            \cE_{1 : i - 1}
        } 
        = \frac{s(j_1)}{ \sum_{j \in \cC_1} s(j) }
        \cdot 
        \frac{s(j_2)}{ \sum_{j \in \cC_2} s(j) }
        \cdots
        \frac{s(j_k)}{ \sum_{j \in \cC_k} s(j) }.
    \end{align}

\end{proof}

\section{Lower Bound for Approximate DP with $\ell_\infty$ neighboring Relationship}
\label{sec:lowerbound}

In this section, we prove the lower bound stated in Theorem~\ref{thm:lower bound of private MST}.  
Our proof consists of two steps: in Section~\ref{sec:reduction-from-epsilon-delta-dp-to-1-delta-dp}, we present a simple reduction from $(\eps, \delta)$-DP MST algorithms to $(1, \delta)$-DP MST algorithms to facilitate the derivation of lower bounds; in \cref{sec:1-delta-dp-lower-bound}, we present a lower bound for all $(1, \delta)$-DP MST algorithms, which can be extended to $(\epsilon, \delta)$-DP lower bound using the previous reduction. 

\subsection{Reduction from $(\eps, \delta)$-DP to $(1, \delta)$-DP}
\label{sec:reduction-from-epsilon-delta-dp-to-1-delta-dp}

\begin{lemma}
    Assume that $\eps < 1$.
    Suppose that there is an $(\eps, \frac{e^{\eps} - 1}{e - 1} \cdot \delta)$-DP MST algorithm $\cM$ such that for every input graph $G = (V, E, \vec{W})$ (with the MST denoted by $T^*$), it holds that  
    \begin{align}
        \begin{array}{c}
            \E{\cM}{ \sum_{e \in \cM(G)} w_e } - \sum_{e \in T^*} w_e 
            \in o \PAREN{ 
                \PAREN{1 / \eps} \cdot n^{3 / 2} \cdot \ln n
            }.
        \end{array}
    \end{align}
    Then there exists a 
    $\PAREN{ 
        1, \delta
    }
    $
    -DP MST algorithm $\cM'$, such that for every input graph $G' = (V', E', W')$ (with the MST denoted by $T'$), it holds that 
    \begin{align}
        \label{ineq: impossible upper bound of 1 delta MST algorithm}
        \begin{array}{c}
            \E{\cM'}{ \sum_{e \in \cM'(G')} w_e' } - \sum_{e \in T'} w_e' 
            \in o \PAREN{ 
                n^{3 / 2} \cdot \ln n
            },
        \end{array}
    \end{align}
\end{lemma}

\begin{proof}
    We demonstrate how to construct $\cM'$ based on $\cM$.  
    Given a graph $G' = (V', E', W')$, we create a new graph $G = (V, E, W)$ as follows: $V = V'$, $E = E'$, and $W = (1 / \eps) \cdot W'$.  
    To simplify the discussion, we assume that $1 / \eps \in \N^+$; otherwise, we replace $1 / \eps$ with $\lceil 1 / \eps \rceil$, which affects the privacy and utility guarantees only by a constant factor, as $\eps \leq 1$.  
    Based on this construction, we have $T' = T^*$.  
    Finally, we define $\cM'$ as:  
    $
        \cM'(G') \doteq \cM(G).
    $
    Then based on group privacy property (Fact~\ref{fact: group privacy}), $\cM'$ is 
    $
        \PAREN{ 
            \eps \cdot \frac{1}{\eps}, \frac{e^{\frac{1}{\eps} \cdot \eps} - 1}{e^{\eps} - 1} \cdot \frac{e^{\eps} - 1}{e - 1} \cdot \delta
        }
        =
        \PAREN{ 
            1, \delta
        }
    $
    -DP.
    We conclude the proof by observing that
    \begin{align}
        \begin{array}{rl}
            \E{\cM'}{ \sum_{e \in \cM'(G') } w_e' } 
                &\hspace{-2mm} = \E{\cM}{
                    \sum_{e \in \cM(G) } w_e'
                } 
                = \E{\cM}{
                    \sum_{e \in \cM(G) } \eps \cdot w_e
                } 
                = \eps \cdot \E{\cM}{
                    \sum_{e \in \cM(G) } w_e
                } \vspace{1.5mm} \\
                &\hspace{-2mm} 
                \in \eps \cdot \PAREN{ \sum_{e \in T^*} w_e + o \PAREN{ 
                        \PAREN{1 / \eps} \cdot n^{3 / 2} \cdot \ln n
                    }
                }
                = \sum_{e \in T'} w_e' + o \PAREN{ 
                        n^{3 / 2} \cdot \ln n
                    }.
        \end{array}
    \end{align}
\end{proof}

\subsection{$(1, \delta)$-DP Lower Bound}
\label{sec:1-delta-dp-lower-bound}

In this subsection, we show that no $(1, \delta)$-DP algorithm $M$ satisfies~\cref{ineq: impossible upper bound of 1 delta MST algorithm} for $\delta = O(1/\sqrt{n})$, thereby proving Theorem~\ref{thm:lower bound of private MST}.  
Our proof heavily relies on a technique originally developed by~\citet*{SteinkeU17} for establishing lower bounds for the private maximum top-$k$ selection problem.  
To apply this technique, we negate the weights of the input graph. 
Instead of proving a lower bound for the minimum spanning tree problem, we demonstrate that any $(1, \delta)$-DP maximum spanning tree algorithm $M$ incurs an error of $\Omega(n^{3/2} \log n)$ for $\delta = O(1/\sqrt{n})$.

\begin{theorem}[Theorem 3 in \citep{SteinkeU17}]
    \label{thm:fingerprint}
    Let $\beta, \gamma, \Delta, k > 0$ and $\numSample, d \in \N^+$ be a fixed set of parameters. 
    Let $P = \PAREN{ P_1, \dots, P_d }$ be independent samples from the beta distribution $\BetaDist{\beta}{\beta}$, and let $X \in \{0, 1\}^{\numSample \times d}$ be a random dataset such that every $X_{i, j}$ is an independent sample from Bernoulli distribution with mean $P_j$ for every $i \in [\numSample]$ and $j \in [d]$. 

    Let $\cA: \{0, 1\}^{\numSample \times d} \to \R^d$ be a $(1, \beta\gamma k / (\numSample \Delta))$-differentially private algorithm (where two datasets $X$ and $X'$ are neighboring if and only if they differ by at most one row). 
    Assume $\E{P, X, \cA}{ \norm{\cA(X)}_2^2 } = k$ and $\P{\norm{\cA(X)}_1 \le \Delta} = 1$ and 
    \begin{equation}
        \begin{array}{c}
            \E{P, X, \cA}{\sum_{j \in [d]} \cA (X)_j \cdot \left(P_j - \frac{1}{2}\right)} \ge \gamma k ~,
        \end{array}
    \end{equation}
    then $\numSample \ge \beta\gamma\sqrt{k}$. 
\end{theorem}

Informally speaking, from \cref{thm:fingerprint}, any DP algorithm with a good utility requires a lot of samples. 
To translate the lower bound on samples to a lower bound on the utility for MST, our goal is to construct a hard distribution of graphs, where the weight vector simulates the random dataset $X$ and the algorithm's utility ties to the same quantity in the theorem. 
The proof is by contradiction. 
When the weight vector, or equivalently, the number of samples, is fixed, \cref{thm:fingerprint} shows that any DP algorithm with a "good" utility requires more samples than this fixed number, leading to a contradiction. 
We will present the proof formally in the rest of this section. 

\begin{definition}[Hard Distribution]
    \label{def:hard-distribution}
    Consider a complete graph $G$ with $n$-vertices and let $m = n (n-1) / 2$ be the number of edges. 
    Let $\beta, \gamma > 0$ and $\numSample \in \N^+$ be parameters to be determined later. 
    For each edge $e$, we first sample $P_e$ from $\BetaDist{\beta}{\beta}$ independently.
    Then let $X \in \{0, 1\}^{\numSample \times m}$ be a random dataset such that $X_{i, e}$ is an independent sample from Bernoulli distribution with mean $P_e$, $\forall i \in [\numSample], e \in [m]$. 
    Finally, let $w_e \doteq \sum_{i \in [\numSample]} X_{i,e}$ for each $e \in [m].$
    Hence, $w_e$ follows binomial distribution $\cB(\numSample, P_e)$.
\end{definition}

Since a spanning tree algorithm $M$ has a different input and output format from the algorithm in Theorem~\ref{thm:fingerprint}, a conversion is required. 

\begin{definition}[Conversion]
    \label{def:alg-conversion}
    Let $G$ be the random graph and $X$ the random dataset generated as described in \cref{def:hard-distribution}.  
    Given an algorithm $M$ that takes $G$ as input and outputs a set of spanning tree edges, define $\cA_M$ as the algorithm that, given input $X$, outputs an indicator vector $\cA_M(X) \in \set{0, 1}^{|E|}$, where $\cA_M(X)_e = 1$ if and only if $e \in M(G)$.  
\end{definition}

It is straightforward to see that if $M$ is $(1, \delta)$-DP (where two input graphs are considered neighboring if the weights of each edge differ by at most $1$), then $\cA_M$ is also $(1, \delta)$-DP (where two input datasets are neighboring if they differ in at most one row).  
Therefore, under the hard distribution, the sample complexity lower bound in \cref{thm:fingerprint} for the algorithm $\cA_M$ transfer to $M$.

Before presenting our main lower bound, we state a lemma that is critical to our proof.  

\begin{lemma}
    \label{claim:random-graph}
    For $\numSample \ge 10$, $\beta = \frac{1}{2}\ln n$ and $n \ge 2 \times 10^7$, let $G$ be a random graph sampled from the hard distribution in \cref{def:hard-distribution}. 
    With probability at least $0.99$, there exists a spanning tree of \ $G$ where each edge has weight at least $3 / 4 \cdot \numSample$. 
\end{lemma}

The complete proof of \cref{claim:random-graph} is included in Appendix~\ref{proof:random-graph}.  
Intuitively, the result follows from the properties of the beta distribution and the Erdős–Rényi random graph model~\citep{erdos59a}.  
Let $H$ be the subgraph of $G$ that includes all vertices and all edges $e$ such that $w_e \geq \frac{3}{4} \cdot \numSample$.  
Since the edge weights are independent and due to our choice of $\beta$, the graph $H$ is an instance of the Erdős–Rényi model with edge sampling probability $p = \P{w_e \geq \numSample \cdot 3/4} \in \Omega \left( \ln(n)/n \right)$.
Finally, $G$ has a spanning tree where all edges $e$ satisfy $w_e \geq \numSample \cdot 3/4$ if and only if $H$ is connected, which occurs almost surely when $p \in \Omega \left( \ln(n)/n \right)$~\citep{erdos59a}.

Assume \cref{claim:random-graph} holds, we directly have
\begin{equation}
    \label{eq:large-mst}
    \begin{array}{c}
        \E{G}{\sum_{e \in T^*} w_e} \ge 0.99 \cdot \frac{3}{4} \cdot \numSample \cdot (n - 1).
    \end{array}
\end{equation}
where $T^*$ be the maximum spanning tree in $G$.
Next, we will present the main theorem in this section and captures the utility lower bound. 

\newcommand{\error}{\frac{n^{3/2}\ln n}{1000}}

\begin{theorem}[MST lower-bound]
    \label{thm:mst lower bound}
    Let $n \ge 9 \times 10^8$, $\beta = \frac{1}{2}\ln n$ and $\numSample = \frac{1}{100}\sqrt{n}\ln n$, $G$ be a random graph 
    sampled from the hard distribution in \cref{def:hard-distribution}. 
    Let $\gamma = 0.04$ and 
    $\delta = \beta\gamma/\numSample = 2 / \sqrt{n}$. 
    For any $\left(1, \delta\right)$-differentially private maximum spanning tree algorithm $M$, it holds that 
    \begin{equation}
        \label{ineq:mst lower bound}
        \begin{array}{c}
            \E{G, M}{\sum_{e \in M(G)} w_e} \le \E{G}{\sum_{e \in T_G^*} w_e} - \error,
        \end{array}
    \end{equation}
    where $T_G^*$ be the maximum spanning tree given $G$.
\end{theorem}
\begin{proof}
    The proof proceeds by contradiction.  
    Assume there exists a $(1, \delta)$-DP maximum spanning tree algorithm $M$, s.t. 
    \vspace{-5mm}
    \begin{equation}
        \begin{array}{c}
            \E{G, M}{\sum_{e \in M(G)} w_e} > \E{G}{\sum_{e \in T_G^*} w_e} - \error,
        \end{array}
    \end{equation}
    Let $X$ be the random dataset generated when $G$ is constructed, as described in \cref{def:hard-distribution}, and let $\cA_M$ denote the algorithm converted from $M$, as described in \cref{def:alg-conversion}.
    We will prove that
    \begin{align}
        \label{ineq: mst lower bound for beta parameter estimation}
        \begin{array}{c}
            \E{P, X, \cA_M}{ \sum_{e \in [m]} \PAREN{ \cA_M(X) }_e (P_e - \frac{1}{2}) } 
                > \gamma (n - 1),
        \end{array}
    \end{align}
    Since $M$ outputs $(n - 1)$ edges, we always have $\norm{\cA_M(X)}_2^2 = \norm{\cA_M(X)}_1 = n - 1$.  
    Hence, $\cA_M$ satisfies all conditions in \cref{thm:fingerprint}, where we set $k = \Delta = (n - 1)$ and $\delta = \beta \gamma k / (\numSample \Delta) = \beta \gamma / \numSample$.  
    Therefore, \cref{thm:fingerprint} implies that $\numSample \ge \beta \gamma \sqrt{n - 1}$, leading to a contradiction since the assumption in \cref{thm:mst lower bound} implies $\numSample = 0.01 \sqrt{n} \ln n < \beta \gamma \sqrt{n - 1}$.

    \vspace{2mm}
    \noindent {\it Proving \cref{ineq: mst lower bound for beta parameter estimation}:}
    Since $w_e \sim \cB(\numSample, P_e)$, we have $\E{w_e} = \numSample \cdot P_e.$
    Further, the following lemma holds, whose proof employs a standard technique using Hoeffding's inequality and is deferred to \cref{proof: maximum of subgaussin}.
    \begin{lemma}
        \label{lemma: maximum of subgaussin}
        $
            \E{ \max_{e \in E} |w_e - \numSample \cdot P_e| } \le \sqrt{3\numSample\ln n}.
        $
    \end{lemma}
    \noindent
    Combing \cref{lemma: maximum of subgaussin} and that $\norm{\cA_M(X)}_1 \equiv n - 1$, we see 
    \begin{equation}
            \hspace{-2mm}
            \E{P, X, \cA_M}{ 
                \hspace{-1.5mm}
                \begin{array}{c} 
                    \sum_{e \in [m]} \PAREN{ \cA_M(X) }_e ( \numSample \cdot P_e - \frac{\numSample}{2}) 
                \end{array}
                \hspace{-1.5mm}
            } 
            \ge \E{P, X, \cA_M}{ 
                \hspace{-1.5mm}
                \begin{array}{c} 
                    \sum_{e \in [m]} \PAREN{ \cA_M(X) }_e w_e 
                \end{array}
                \hspace{-1.5mm}
            } 
            \hspace{-1.5mm}
            \begin{array}{c} 
                - (n - 1) \paren{ \sqrt{3 \numSample\ln n} + \frac{\numSample}{2} }
            \end{array}
            \hspace{-1.5mm}
    \end{equation}
    By the construction of $\cA_M$ (\cref{def:alg-conversion}), along with the assumption that $M$ violates \cref{ineq:mst lower bound}, 
    
    \begin{equation}
            \E{P, X, \cA_M}{ 
                \hspace{-1.5mm}
                \begin{array}{c} 
                    \sum_{e \in [m] } \PAREN{ \cA_M (X) }_e w_e 
                \end{array}
                \hspace{-1.5mm}
            } 
            = \E{G, M}{ 
                \hspace{-1.5mm}
                \begin{array}{c} 
                    \sum_{e \in M(G) } w_e 
                \end{array}
                \hspace{-1.5mm}
            } 
            > \E{G}{ 
                \hspace{-1.5mm}
                \begin{array}{c} 
                    \sum_{e \in T_G^* } w_e
                \end{array}
                \hspace{-1.5mm}
            } 
            - \hspace{-1.5mm}
                \begin{array}{c} 
                    \error
                \end{array}
            \hspace{-1.5mm}.
    \end{equation}

    \noindent Hence
    \begin{align*}
        \E{P, X, \cA_M}{ 
                \hspace{-1.5mm}
                \begin{array}{c} 
                    \sum_{e \in [m]} \PAREN{ \cA_M(X) }_e ( \numSample \cdot P_e - \frac{\numSample}{2}) 
                \end{array}
                \hspace{-1.5mm}
            }  
            &> \E{G}{ 
                \hspace{-1.5mm}
                \begin{array}{c} 
                    \sum_{e \in T_G^* } w_e
                \end{array}
                \hspace{-1.5mm}
            } 
            \hspace{-1.5mm}
                \begin{array}{c} 
                    - \error - (n - 1) (\sqrt{3\numSample  \ln n} + \frac{\numSample}{2}) 
                \end{array}
            \hspace{-1.5mm} \\
            &> 0.99 \cdot \frac{3}{4}\numSample(n - 1) 
            \hspace{-1.5mm}
                \begin{array}{c} 
                    - \error - (n - 1) (\sqrt{3\numSample  \ln n} + \frac{\numSample}{2}) 
                \end{array}
            \hspace{-1.5mm} \\
            &= 0.2425 \numSample (n - 1) 
            \hspace{-1.5mm}
                \begin{array}{c} 
                    - \error - (n - 1)\sqrt{3\numSample\ln n}
                \end{array}
            \hspace{-1.5mm}.
    \end{align*}

    \noindent
    Since $n \ge 9 \times 10^8, \numSample  = \frac{1}{100} \sqrt{n} \ln n$ and $\gamma = 0.04$, it follows that 
    \begin{align*}
            \E{P, X, \cA_M}{ 
                \hspace{-1.5mm}
                \begin{array}{c} 
                    \sum_{e \in [m]} \PAREN{ \cA_M(X) }_e ( P_e - \frac{1}{2}) 
                \end{array}
                \hspace{-1.5mm}
            } 
            \begin{array}{c}
                > 0.2424(n - 1) - \frac{1}{\numSample}\left((n - 1)\sqrt{3\numSample\ln n} + \error\right)  
                > \gamma(n - 1) ~.
            \end{array}
    \end{align*}
\end{proof}

\section{Related Work}
\label{sec:summary}

Releasing various graph statistics under differential privacy constraints is a fundamental and well-studied task. 
For more background, see \cite{li2023survey,mueller2022sok}.

\subsection{Top-$k$ Selection.}\label{sec:top-k}

The private MST problem is closely related to the private top-$k$ selection problem. 
Formulated in the framework of this paper, the problem involves $m$ items with weights $w_1, \ldots, w_m$, and the goal of private top-$k$ selection is to privately sample $k$ items with approximately maximum or minimum weights.
Notably, maximization and minimization are equivalent by flipping the sign of all weights. 
The private MST problem can be seen as a variant of top-$(n - 1)$ selection with the additional topological constraint that the selected items must form a spanning tree.

\citet*{darfee2019} introduced the first linear-time algorithm for private top-$k$ selection with asymptotically optimal privacy-utility trade-off under approximate DP.
They showed that iteratively applying exponential mechanism \citep{mcsherry2017} to select $k$ items with approximately maximum weights, is equivalent to adding random noises following Gumbel distribution (see \cref{def:gumbel-distribution}) to the item weights, and then returning the $k$ items with maximum noisy weights. 
This technique has a rich research history in the context of the non-private top-$k$ problem~\citep{YELLOTT1977, ohlsson1990sequential, ROSEN1997135, Cohen97}, appearing under various names. 
For the top-$1$ maximum selection problem, \citet{YELLOTT1977} showed very early that adding Gumbel noise to the $(\ln w_i)$'s and selecting the maximum allows sampling an item with probability proportional to their weights. 
\citet{ROSEN1997135} studied the top-$k$ maximum selection problem under the name of "order sampling" and proposed a method to sample $k$ items with probability proportional to their weights, by generating noisy scores $\ExpNoise{w_i} = \ExpNoise{1} / w_i$ for $i \in [m]$ and selecting $k$ items with the minimum scores. 
This is equivalent to finding the $k$ items maximizing $-\ln \ExpNoise{1} + \ln w_i$, where $- \ln \ExpNoise{1}$ follows exactly a Gumbel distribution (see \cref{fact:Exp and Gumbel}).

Our sampling model, PPSACR, and the corresponding technique in \cref{alg:iterative}, extend this line of research~\citep{Cohen97, YELLOTT1977, ROSEN1997135} by enabling the adaptive removal of candidate items after each sampling step, based on previously selected items.

\citet{QiaoSZ21} proposed a linear-time private top-$k$ selection algorithm that achieves an asymptotically optimal privacy-utility trade-off (up to a logarithmic factor) under approximate differential privacy. 
Their approach involves adding independent Laplace noise to each item weight and returning the set of top-$k$ items with the smallest noisy weights.  
However, their current privacy-utility trade-off analysis relies heavily depends on the returned items corresponding to the true $k$ smallest noisy weights. 
Extending their approach and analysis to the private MST problem remains open, as the selected items or edges may not necessarily form a spanning tree.

\subsection{Releasing an MST under DP}

We supplement results for \emph{approximate DP} (\cref{tab:summ}) with results for \emph{pure DP} in \cref{tab:summ-pure}.
Before our work there were two approaches to private MST: \emph{input privatization} and \emph{in-place} algorithms \cite{Pinot_2018} that we review next. 

\begin{table*}[t]
\centering
\resizebox{0.9\columnwidth}{!}{%
    \begin{tabular}{l|c|c|c|c|l}
    {\bf \newline Reference} & {\bf PN }& {\bf NH} & {\bf Error} & {\bf Time}  & {\bf Technique}\\
    \hline
    \cite{Sealfon_2016} & $\epsilon$-DP & $\ell_1$ & $O\PAREN{\paren{1/\epsilon}\cdot  n \log n}$ & MST + $O(m)$ & Input privatization\\
    \cite{pinot_2018_ma} & $\epsilon$-DP & $\ell_1$ & $O\PAREN{\paren{1/\epsilon}\cdot n \log n}$ &  $O(n m)$ &  In-place noise\\
    \cite{hladik_tetek_2024} & $\epsilon$-DP & $\ell_1$  & $\Omega \PAREN{\paren{1/\epsilon}\cdot n\log n} $ & -- & \textit{Lower bound}\\
    \hline
    \cite{Sealfon_2016} & $(\epsilon, \delta)$-DP & $\ell_1$ & $O \PAREN{\paren{1/\epsilon}\cdot n \sqrt{\paren{\log n} \cdot \log \paren{1 / \delta}}}$ & MST + $O(m)$ & Input privatization\\
    \cite{pinot_2018_ma} & $(\epsilon, \delta)$-DP & $\ell_1$ & $O\PAREN{\paren{1/\epsilon}\cdot n \sqrt{ \paren{\log n} \cdot \log \paren{1 / \delta} )}}$ &  $O(n m)$ &  In-place noise\\
    \cite{Sealfon_2016} & $(\epsilon, \delta)$-DP & $\ell_1$  & $\Omega(n) $ & -- & \textit{Lower bound}\\
    \hline
    \cite{Sealfon_2016} & $\epsilon$-DP & $\ell_\infty$  & $O\PAREN{\paren{1/\epsilon}\cdot nm \log n}$ & MST + $O(m)$ & Input privatization\\
    \cite{pinot_2018_ma} & $\epsilon$-DP & $\ell_\infty$ & $O\PAREN{\paren{1/\epsilon}\cdot n^2 \log n}$ & $O(nm)$ & In-place noise\\
    \cite{hladik_tetek_2024} & $\epsilon$-DP & $\ell_\infty$  & $\Omega\PAREN{\paren{1/\epsilon}\cdot  n^2\log n} $ & -- & \textit{Lower bound}\\
    \hline
    \end{tabular}
}\\
\caption{Extended landscape of results and complementing \cref{tab:summ} with further known results for a graph with $n$ vertices and $m$ edges.
\textbf{PN} and \textbf{NH} denote the \emph{Privacy Notation} and the \emph{Neighboring Relationship} respectively.
The table compares the previous works \cite{Sealfon_2016, pinot_2018_ma,hladik_tetek_2024} for both the $\ell_1$ and $\ell_\infty$ neighborhood relation.
``MST'' in the {\bf Time} column is the running time of any non-private MST algorithm.
}
\label{tab:summ-pure}
\vspace{-6mm}
\end{table*}

\paragraph{In-place}

Both Prim-Jarník's and Kruskal's algorithm \cite{kruskal_1956,Prim1957,jarnik1930} start with an empty set of edges and then iteratively grow it while guaranteeing that it is still a subset of an MST.
In each step, they greedily select the lightest new edge between cuts that respect the edges already chosen.
One can privatize them by injecting noise whenever a weight is accessed during the computation of an MST.
We can replace the selection step using any differentially private selection mechanism, for instance, \emph{Report-Noisy-Max}~\cite{dwork_roth_2014}, \emph{Permute-and-Flip}~\cite{mckenna_sheldon_2020}, or the \emph{Exponential Mechanism}~\cite{mcsherry2017} and get overall privacy by composition.
PAMST \cite{pinot_2018_ma,hladik_tetek_2024} is based on the Prim-Jarník algorithm and gives the same asymptotic utility as our approach.
Furthermore, a private version of Kruskal's algorithm has been suggested by \citet{McKenna_Miklau_Sheldon_2021} as a subroutine in privately generating synthetic data using a probabilistic graphical model, but its utility has never been discussed.
We provide such an analysis in \cref{sec:privateKruskal}.
Compared to the \emph{input privatization} in the next paragraph \cite{Sealfon_2016}, this approach gives strictly better utility under the $\ell_\infty$ neighboring relationship, assuming the graph is not too sparse.
Unfortunately, the caveat is the running time because of the cost of the noisy selection.
In an unpublished manuscript \cite{pagh2024fasterprivateminimumspanning}, a subset of this paper's authors brought down the running time to $O(m + n^{3/2} \log(n) / \sqrt{\rho})$  by designing a special priority queue for Prim's algorithm, which simulates \emph{Report-Noisy-Max} in sublinear time $O(\sqrt{n/\rho} \log n)$, where $\rho$ can be chosen as in Algorithm~\ref{algo:impl-details} to ensure $(\eps, \delta)$-DP.

\vspace{-1mm}
\paragraph{Input privatization} 
One simple idea is to release a private synthetic graph by adding noise to all the edge weights and obtain privacy by post-processing.
\citet{Sealfon_2016} was the first to analyze \emph{input perturbation} using Laplace noise to achieve $\varepsilon$-DP under the $\ell_1$ neighborhood and gave error of $O(n \log n)$ which is known to be asymptotically optimal \cite{hladik_tetek_2024}.
However, under $\eps$-DP with $\ell_\infty$ neighboring relationship, this technique gives an additive error of $O(n m \log n)$.
For dense graphs, this can leave a gap of up to a factor of $O(n)$ to the known lower bound of $\Omega(n^2 \log n)$ under $\eps$-DP.
The advantage is that it allows flexibility in choosing any (non-private) MST algorithm e.g. the expected linear time algorithm by \citet{karger1996}, the deterministic near-linear time algorithm by \citet{chazelle2000}, or the deterministic linear time for dense graphs by \citet{fredman1987}.
The overall running time is compounded by the $O(m)$ time it takes to add fresh noise to each edge and the running time of the chosen MST algorithm.
Another advantage is that releasing a single private synthetic graph simultaneously allows the computation of other graph statistics, such as finding \emph{shortest paths} or \emph{minimum weight perfect matchings} \cite{Sealfon_2016}.

\paragraph{Lower Bounds} 
Recent work by \citet{hladik_tetek_2024} showed tight asymptotical worst-case bounds for $\epsilon$-DP using a packing argument: $\Omega(n \log n/\epsilon)$ for the $\ell_1$ neighborhood and $\Omega(n^2 \log n/\epsilon)$ for $\ell_\infty$.
They improved \citeauthor{Sealfon_2016}'s $\ell_1$ bound of $\Omega(n)$.
Under the $\ell_\infty$ neighborhood, we show in \cref{th:upper-bound} together with \cref{thm:lower bound of private MST}, that an expected error of $\tilde{\Theta}(n^{3/2})$ is asymptotically tight.

\section{Empirical Evaluation}\label{sec:empiricalevaluation}

To confirm our theoretical claims, we implemented PAMST \cite{pinot_2018_ma}, Sealfon's input privatization \cite{Sealfon_2016}, and our \cref{algo:impl-details} (instantiated with Prim-Jarník) in \emph{Python 3.9}.
The implementation relies on version $3.2.1$ of the \emph{NetworkX} library \cite{networkx2008}.
All experiments ran locally on a MacBook Pro with an Apple M2 Pro processor (10 Cores, up to 3.7GHz) and 16GB of RAM.
The first experiment resembles a natural setting in the context of synthetic data generation, and the second explores the influence of the graph's density.
The experiments indicate that the output distribution of our algorithm indeed matches PAMST and, hence, outperforms the \emph{input privatization} approach if the graph is not too sparse.
The results are shown in \cref{appx:experiments}.

\section{Conclusion and Open Problems}

Our work shows that a simple input perturbation yields a privacy guarantee for the output of any MST algorithm far better than what is implied by the post-processing property of DP.
It is natural to wonder if something similar holds for other problems where the best existing in-place private algorithms add noise during the computation.

There remains a small gap of $\mathcal{O}\paren{\sqrt{\log(1/\delta)}}$ between known upper and lower utility bounds.
As discussed in \cref{sec:top-k}, extending the Laplace noise-based approach for top-$k$ selection by \citet{QiaoSZ21} to the private MST problem remains an open question.

\begin{acks}
Pagh, Wu, Zhang, and Retschmeier carried out this work at Basic Algorithms Research Copenhagen (BARC), which was supported by the VILLUM Foundation grant 54451. 
\emph{Providentia}, a Data Science Distinguished Investigator grant from the Novo Nordisk Fonden, supported Pagh, Retschmeier, and Wu.
Hanwen Zhang is also partially supported by Starting Grant 1054-00032B from the Independent Research Fund Denmark under the Sapere Aude research career programme.
We thank Edith Cohen for providing us with historical context for PPSACR and anonymous reviewers of a previous version of this paper for their valuable feedback.
\end{acks}

\bibliographystyle{ACM-Reference-Format}
\bibliography{references.bib}
\appendix
\newpage

\section{Algorithm for Maximum Weight Independent Set in a Matroid}
\label{appendix:matroid}

A matroid $(U, \cI)$ is defined by a finite ground set $U$ and a family of \emph{independent sets} $\cI \subseteq 2^U$. 
The family $\cI$ satisfies the following properties:
\begin{itemize}
    \item $\emptyset \in \cI$; 
    \item $\forall S \in \cI$ and $T \subseteq S$, $T \in \cI$; 
    \item $\forall S, T \in \cI$ and $|S| < |T|$, there exists $t \in T \setminus S$ such that $S \cup \{t\} \in \cI$. 
\end{itemize}
As an example, let $U$ be the set of edges in a graph, and $\cI$ be the family of all subsets of edges that form a forest. 
It's easy to verify that $(U, \cI)$ indeed is a matroid, and this type of matroid is usually called a \emph{graphic matroid}. 
Given a weight function $w: U \to \R^+$, the weight of an independent set $S$, $w(S)$, is defined as the sum of the weight of its elements. 
Therefore, the maximum spanning tree problem (and hence the MST problem) is a special case of finding a maximum weight independent set in this matroid. 
Kruskal's algorithm for finding a maximum spanning tree was generalized to the problem of finding a maximum weight independent set in a matroid by \citet{edmonds1971matroids}. 
The algorithm first sorts all the elements in the matroid by weight in decreasing order and then tries to insert elements one by one into the independent set if possible. 
Since the analysis for \cref{algo:impl-details} is based on Kruskal's algorithm, it naturally generalizes to finding a maximum weight independent set in general matroids. 
For matroids with rank $n$ and $m$ elements, since there is no upper bound for $m$ on $n$, the expected error we have will be $O(n^{3/2}\log m/\sqrt{\rho})$. 

\begin{algorithm}[H]
    \caption{\privateMSTFramework: Private Maximum-Weight-Independent-Set-in-Matroid Framework }
    \label{algo:impl-details-matroid}
    \begin{algorithmic}[1]
        \Statex \hspace{-4.5mm} {\bf Input: } a matroid $(U, \cI)$ and the weight $\vec{W}$, any algorithm $\cA_{MWIS(matroid)}$, privacy parameters $\PAREN{\eps, \delta}$
        \State $\rho \leftarrow \bigparen{
            \sqrt{\eps + \log \paren{1 / \delta}} - \sqrt{\log \paren{1 / \delta} \, }
        }^2$;\, 
        $
            \eps' \leftarrow \sqrt{{2 \cdot \rho} / \, \paren{n - 1}}
        $;\, 
        \State $\tilde{w}_e \leftarrow w_e + \PAREN{2 / \, \eps'} \cdot \ln \left(\ExpNoise{1}\right)$ for all $e \in U$
        \State \textbf{return} $\cA_{MWIS(matroid)}(U, \cI, \tilde{\vec{W}})$
    \end{algorithmic}
\end{algorithm}

\section{One-Shot-PPSACR}

\label{appendix: One-Shot-PPSACR}
\begin{algorithm}[H]
    \caption{One-Shot-PPSACR}
    \label{alg:one-shot-PPSACR}
    \begin{algorithmic}[1]
        \State $\Tilde{s}(j) \leftrightarrow \ExpNoise{1} / s(j), \forall j \in U.$
        \State $\cC \leftarrow U, \, \cI \leftarrow$ An Empty List \Comment{possible candidates and sampled items so far}
        \While{$|\cI| \le k$ and $\cC \neq \varnothing$}
            \State $j \leftarrow \argmin_{j' \in \cC} \Tilde{s}(j)$.
            \State Add $j$ to the back of $\cI$
            \State $\cC \gets \cC \setminus \PAREN{ \set{ j } \cup f \big( \cI \big) }$
        \EndWhile
        \State \textbf{return} $\cI$
    \end{algorithmic}
\end{algorithm}

\newpage

\section{Probabilities}
\label{appendix: proofs for premilinary}

\noindent
The distribution of the logarithm of an exponential random variable is closely connected to the \emph{Gumbel distribution.}

\begin{definition}[Gumbel Distribution]
    \label{def:gumbel-distribution}
    Given parameter~$b \in \R$, the Gumbel distribution,~$\GumbelNoise{b}$, has probability density function $
        p(z) = \frac{1}{b} \cdot \exp \PAREN{ - \PAREN{ \frac{z}{b} + \exp \PAREN{ - \frac{z}{b} } }  },
    $ 
    $\forall z \in \R$, and cumulative distribution function $
        F(z;  b) = \exp \paren{- e^{-z / b }}, \forall z \in \R.
    $
\end{definition}

\begin{fact}
    \label{fact:Exp and Gumbel}
    If $Z \sim \ExpNoise{1}$, then $-\ln Z \sim \GumbelNoise{1}$.
\end{fact}

\begin{proof}[Proof of \cref{fact:Exp and Gumbel}]
    \begin{equation}
        \P{-\ln X \le z}
            = \P{X \le \exp \PAREN{-z}}
            = \exp \PAREN{ - e^{-z} },
        \quad \forall z \in \R.
    \end{equation}
\end{proof}

\vspace{3mm}
\noindent {\bf Proof of Fact~\ref{fact: confidence interval of gumbel noise}.}
    For our proof, we require the following fact, which we will prove later.

    \begin{fact}\label{fact: confidence interval of gumbel noise proof}
        If $Z_i \sim \ln \ExpNoise{1}$, then for each $\beta \in (0, 1)$, 
        \begin{align*}
            \P{ Z_i > \ln \ln \frac{1}{\beta} } 
                = 
            \P{ Z_i < \ln \ln \frac{1}{1 - \beta} } 
                = \beta.
        \end{align*}
    \end{fact}

    \noindent
    Since $\ln (1 + x) \le x$, $\forall x > -1$, it holds that $\ln \ln \frac{1}{1 - \beta} = \ln \paren{ - \ln \paren{1 - \beta} } \ge \ln \beta = - \ln \frac{1}{\beta}$.
    It follows that
    \begin{align*}
        \P{ Z_i < - \ln \frac{1}{\beta} }
            \le \P{ Z_i > \ln \ln \frac{1}{\beta} }     
            = \beta.
    \end{align*}    
    Define $X \doteq \max_{i \in [m]} |Z_i|$. 
    By union bound, we have 
    \begin{align*}
        \P{ \card{X} \ge \ln \frac{2 \cdot m}{\beta} } \le \beta,
        \qquad  \forall \beta \in \paren{0, 1}.
    \end{align*}
    Define $t \doteq \ln \frac{2 \cdot m}{\beta} \ge \ln \paren{ 2 \cdot m}$.
    Then $\beta = \paren{2 \cdot m} \cdot e^{-t}$.
    Therefore, 
    \begin{align}
        \E{\card{X}}
            &= \int_0^{\ln (2 \cdot m)} \P{ \card{X} \ge t } \, d t + \int_{\ln (2 \cdot m)}^\infty \P{ \card{X} \ge t } \, d t  \\
            &\le \ln (2 \cdot m) + \int_{\ln (2 \cdot m)}^\infty \paren{2 \cdot m} \cdot e^{-t} \, d t  \\
            &\le \ln (2 \cdot m) + \paren{2 \cdot m} \cdot e^{ - \ln (2 \cdot m) } \\
            &= \ln (2 e \cdot m).
    \end{align}

\begin{proof}[Proof of \cref{fact: confidence interval of gumbel noise proof}]
    If $Y \sim \ExpNoise{1}$ where $\P{Y \leq y}= e^{-y}$ we have for each $\beta \in (0, 1)$, 
    \begin{align*}
        \P{ Y > \ln \frac{1}{\beta} } 
            = \exp \PAREN{ - \ln \frac{1}{\beta} }    
            = \beta, 
        \, \, \text{ and } \,     
        \P{ Y < \ln \frac{1}{1 - \beta} }
            = 1 - \exp \PAREN{ -\ln \frac{1}{1 - \beta} }
            = \beta.
    \end{align*}    
    It follows that 
    \begin{align*}
        \P{ \ln Y > \ln \ln \frac{1}{\beta} } 
            = \P{ \ln Y < \ln \ln \frac{1}{1 - \beta} }
            = \beta.
    \end{align*}    
\end{proof}

\vspace{3mm}
\noindent {\bf Proof of Fact~\ref{fact: scaled exponential}.}
    \begin{align*}
        p_{Y} (y) 
            = p_X \PAREN{ \lambda \cdot y } \cdot \frac{dX}{dY} 
            = \lambda \cdot e^{- \lambda \cdot y},\, 
            \forall \, y \ge 0.
    \end{align*}
\hfill $\square$

\begin{fact}[Two Variables]
    \label{fact:min of two exponential}
    If $X \sim \ExpNoise{\lambda_1}$, and $Y \sim \ExpNoise{\lambda_2}$, then 
    $$ 
        \P{X \le Y}
            = \int_0^\infty \lambda_1 e^{-\lambda_1 x} \cdot \P{Y \ge x} \, dx
            = \int_0^\infty \lambda_1 e^{-\lambda_1 x} e^{-\lambda_2 x} \, dx
            = \frac{\lambda_1}{\lambda_1 + \lambda_2}.
    $$
\end{fact}

\vspace{3mm}
\noindent {\bf Proof of Fact~\ref{fact:min of exps}.}
    Let $Y \doteq \min_{j \in [d] \setminus \set{i}} X_j$.
    Then
    \begin{align}
        \P{ Y \ge y } 
            = \prod_{j \in [d] \setminus \set{i}} \P{ X_j \ge y } 
            = \exp \PAREN{ - y \sum_{j \in [d] \setminus \set{i}} \lambda_j }.
    \end{align}
    Therefore, $Y \sim \ExpNoise{\sum_{j \in [d] \setminus \set{i}} \lambda_j}$.
    Applying \cref{fact:min of two exponential} gives 
    \begin{align}
        \P{ X_i = \min_{j \in [d]} X_{j}} 
            = \frac{\lambda_i}{\sum_{j \in [d]} \lambda_j}
    \end{align}
\hfill $\square$

\vspace{3mm}
\noindent {\bf Proof of Fact~\ref{fact:memoryless of exps}.}
    \begin{align}
        \P{X \ge x + y \mid X \ge x}
            = \frac{\exp\PAREN{x + y}}{\exp\PAREN{x}}
            = \exp\PAREN{y}
            = \P{X \ge y}. 
    \end{align}        
\hfill $\square$

\section{Utility Guarantee of Private Kruskal}
\label{appx:private-kruskal-utility}
We prove the utility of private Kruskal's algorithm (Algorithm~\ref{alg:privKruskal}) directly.
Suppose we use Kruskal's algorithm to find the MST on the noisy and original graphs in parallel. 
When we try to add the $k$-th edge in noisy graph, we are actually finding the noisy minimum-weighted edge among all the cycle-free candidates. 
Also, there must be a candidate from the first $k$ added edges when we run Kruskal's algorithm in the original graph, as all forests form a matroid. 
Therefore, the real minimum weight is not greater than the payment for the $k$-th step on the original graph, so the extra cost induced in this step only comes from selecting the noisy-min. 
Accumulate the error over all steps, we show the desired utility guarantee of the algorithm. 

\begin{proof}\label{proof:utilityPrivateKruskal}

    For each $i \in \{1, \cdots, n-1\}$, let $e_i$ be the edge added to $T$ at the $i^{\text{th}}$ iteration of Algorithm~\ref{alg:privKruskal}.
    Let $T_i \doteq \bigcup_{j \in [i]} \{e_j\}$ be the partial spanning forest after iteration $i$ and for convenience, let $T_0 = \varnothing$.
    Further, initially set $F_0 = E$, and denote with $F_i$ the set $F$ immediately after the $i^{\text{th}}$ iteration containing all edges that still could be added to $T$ without creating a cycle.
    Assume that also run a non-private Kruskal's algorithm on the same graph with the true edge weights in parallel, to obtain the \emph{real MST} $T_i^* = \bigcup_{j \in [i]}\{ e_j^* \}$, where $e_j^*$ is the edge added at iteration $j$.

    To prove the utility guarantee, we want to show that for all steps, with probability $1 - \beta$ and for some universal constant $c \in \R^+$,
    \begin{equation}
        w_{e_i} \le w_{e^*_i} + \frac{c}{\eps'} \cdot \ln \frac{n}{\beta}.
    \end{equation}
    
    \noindent Note that the utility guarantee of the exponetial mechanism gives us with probability $1- \beta$
    \begin{equation}
        w_{e_i} \le \min_{e \in F_{i - 1}} w_{e} + \frac{c}{\eps'} \cdot \ln \frac{n}{\beta}
    \end{equation}
    because $e_i$ is exactly the noisy minimum of the set $F_{i-1}$ chosen by the algorithm.
    Therefore, it suffices to show that 
    \begin{equation}
        \label{eq:private-kruskal-candidate-set}
        \min_{e \in F_{i - 1}} w_{e} \le w_{e_i^*}.
    \end{equation}
    Note that $w_{e_{i-1}^*} \leq w_{e_i^*}$ by the way Kruskal adds new edges to $T^*$.
    Using this, we claim that $F_{i - 1}$ contains at least one edge which belongs to $T_i^*$ and can be added to $T_{i - 1}$ without forming a cycle. 

    First, note that edges in $T_{i - 1}$ induce a forest.
    Let $\cT$ be an arbitrary tree in this forest, and assume that it has $t$ edges. 
    Then, the number of edges $e = (u, v) \in T_i^*$, such that both $u, v \in \cT$ is at most $t$ without inducing a cycle.

    There are $i - 1$ edges in $T_{i - 1}$, but $i$ edges in $T_i^*$.
    Therefore, there must be \emph{at least one} edge from $T^*$, which connects two trees in the forest induced by $T_{i - 1}$.
    It follows that this edge belongs to $F_{i - 1}$, proving \cref{eq:private-kruskal-candidate-set}.

    Summing over all possible edges gives
    $$
        \sum_{i \in [n - 1 ]} w_{e_i} \leq \sum_{i \in [n - 1]} w_{e_i^*} + \frac{c (n - 1)}{\eps'} \ln \frac{n}{\beta}.
    $$
    By a standard integration technique, this also implies an expected error of $O \PAREN{ \frac{n}{\eps'} \ln n }.$
\end{proof}

\section{Efficiently Implementing \privateKruskal}
\label{appx:implPrivKruskal}

This section describes how \privateKruskal can efficiently be implemented.
Although this result is, of course, overshadowed by our main result, we believe it is interesting enough to be stated here.

The algorithm can be implemented in $O(n + m \log n)$ time. 
In the initialization phase, we construct a complete binary tree where every level, except possibly the last, is fully filled, and all nodes in the last level are as far left as possible—with $m$ leaf nodes. 
Each leaf node represents an edge and is assigned a weight $e^{-\eps' w_e}$. 
Each internal node of the binary tree is assigned a weight equal to the sum of the weights of the leaf nodes in the subtree rooted at that node.

This binary tree is used to manage the nodes in $F$ efficiently: it supports both the sampling operation (Algorithm~\ref{alg:privKruskal}, Line~\ref{line: edge sampling}) and the update operation (Algorithm~\ref{alg:privKruskal}, Line~\ref{line: private kruskal end}) in $O(\log n)$ amortized time per edge.

\textit{Sampling.} 
We apply a top-down approach starting from the tree's root to sample an edge (i.e., a leaf node).  
At each step, we move to the left child with probability equal to the ratio of the left child's weight to the current node’s weight, and move to the right child otherwise.  
The sampling procedure terminates when the current node is a leaf.
It is straightforward to verify via an induction on the number of levels that this method samples a leaf node with probability proportional to $e^{-\eps' w_e}$ and that the sampling process completes in $O(\log n)$ time.

\textit{Update.} 
We introduce two auxiliary data structures for this step.  
First, we maintain a union-find data structure \cite{tarjan1975}, which allows us to determine the connected component (induced by the edges in $T$) to which a vertex belongs in $O(\alpha(n))$ amortized time, where $\alpha(n)$ is the inverse Ackermann function.  
Additionally, for each connected component, we maintain a linked list to track the vertices it contains.
When a new edge $e = (u, v)$ is added to the set $T$, let $C_u$ and $C_v$ denote the connected components containing $u$ and $v$, respectively.  
To merge $C_u$ and $C_v$, we update both the union-find structure and the linked list, which can be performed in $O(\alpha(n) + 1)$ time.  
Next, we need to remove from $F$ all edges connecting $C_u$ and $C_v$.  
To do so efficiently, we choose the smaller of $C_u$ and $C_v$ (in term of the number of vertices)--assume, without loss of generality, it is $C_u$--and check all edges incident to it.  
For each edge, if its other endpoint belongs to $C_v$, we remove it from $F$.

We now analyze the number of times an edge is checked and the cost of its removal from $F$.  
For the former, define the rank of a vertex as the number of vertices in the connected component to which it belongs.  
Each time an edge is checked, the rank of one of its endpoints at least doubles.  
Since a vertex can have a rank of at most $n$, an edge can be checked at most $O(\log n)$ times.

For the latter, removing an edge from $F$ is handled by removing the corresponding leaf node from the binary tree and then updating the weights of the nodes along the leaf-to-root path.  
This operation can be completed in $O(\log n)$ time.

\section{Proof of \cref{claim:random-graph}}
\label{proof:random-graph}

Our proof relies on a standard result from the Erdős–Rényi model $\cG(n, p)$, where $\cG$ is a random graph with $n$ vertices such that each edge is independently included with probability $p$.  
The following lemma is stated with explicit constants, as our subsequent proof relies on them.  
Despite extensive searching, we could not find a version of the lemma with explicit constants.  
Therefore, we provide a proof of this lemma with explicit constants at the end of this section.  
Although the proof is standard, we include it for completeness.  

\begin{lemma}[\cite{erdos59a}]\label{lem:erconnectivity}
    For any real $\epsilon > 0$, if $p > \frac{\paren{1 + \epsilon} \ln n}{n}$, then a random graph in the Erdős–Rényi model $\cG(n, p)$ with $n$ vertices is almost surely connected.
    In particular, if $p \ge \frac{2 \cdot \ln n}{n - 1}$ and $n \ge 1700$, the probability that the random graph is not connected is at most $\frac{6}{n}$.
\end{lemma}

Recall that the random graph $G$ considered in \cref{claim:random-graph} is generated as follows: $G$ consists of $n$ vertices and $m = \frac{n (n - 1)}{2}$ edges.
For each edge $e$, we first sample $P_e \sim \BetaDist{\beta}{\beta}$ and then sample $w_e \sim \cB(\numSample, P_e)$. 

Let $H$ be the subgraph which includes all vertices in $G$, and all edges $e$ such that $w_e \ge 3 / 4 \cdot \numSample$. 
Since the edge weights are independent of each other, the graph $H$ is an instance of Erdős–Rényi model with $p = \P{w_e \ge 3 / 4 \cdot \numSample}$. 
Note that $G$ has a spanning tree where all edges $e$ have $w_e \ge 3 / 4 \cdot \numSample$ if and only if $H$ is connected. 
Given \cref{lem:erconnectivity}, to prove \cref{claim:random-graph}, it just remain to show that for each edge $e$, 
\begin{equation}
    \label{eq:edge-probability}
    \P{
        w_e \ge \frac{3}{4} \cdot \numSample
    } 
    > \frac{2 \ln n}{n - 1}. 
\end{equation}
\begin{fact}[\citep{SteinkeU17}]
    \label{fact:beta-distribution-tail}
    Let $\beta \ge 1$, and $X$ be a random variable following beta distribution $\BetaDist{\beta}{\beta}$.
    Then for all $p \in [0, 1 / 2]$, it holds that 
    \begin{align}
        \P{X \le p } = \P{X \ge 1 - p} \ge \PAREN{4 \cdot p \cdot (1 - p)}^{\beta - 1} \cdot \frac{p}{\beta}
    \end{align}
\end{fact}
Firstly, according to \cref{fact:beta-distribution-tail}, 
\begin{align}
    \Pr\left[P_e \ge 0.9\right] \ge \frac{0.36^{\beta - 1}}{10\beta} ~.
\end{align}
Since $\E{w_e} = \numSample \cdot P_e$, by Hoeffding's inequality, 
\begin{align}
    \P{ w_e > \numSample \cdot P_e - 0.15 \numSample \mid P_e \ge 0.9 } 
        < \exp(-0.045 \numSample).
\end{align}
Therefore, 
\begin{align*}
    \P{
        w_e \ge \frac{3}{4}\numSample
    } 
    &\ge \P{P_e \ge 0.9} \cdot \P{w_e \le \numSample\cdot P_e - 0.15\numSample \mid P_e \ge 0.9} \\
    & \ge \frac{0.36^{\beta - 1}}{10\beta}\cdot(1 - \exp(-0.045\numSample)) ~.
\end{align*}
For $\numSample \ge 10$ and $\beta = \frac{1}{2}\ln n$ and $n \ge 2 \times 10^7$, we have
\begin{align}
    \P{
        w_e \ge \frac{3}{4}
    } 
    \ge \frac{0.36^{\frac{1}{2}\ln n}}{1.8 \ln n} \cdot 
    (1 - \exp(-0.45)) 
    > \frac{0.36}{1.8n^{0.52} \ln n} 
    > \frac{2\ln n}{n - 1} ~.
\end{align}

\hfill $\square$

\begin{proof}[Proof of \cref{lem:erconnectivity}]
We need two important lemmas to prove \cref{lem:erconnectivity}.

\begin{lemma}
    \label{lem:x1-upper-bound}
    When $p \ge \frac{2 \cdot \ln n}{n - 1}$, the expected number of connected components of size $1$ in a Erdős–Rényi model $\cG(n, p)$, denoted by $\E{x_1}$, is bounded by
    \begin{equation}
        \E{x_1} \le \frac{1}{n}
    \end{equation}
\end{lemma}
\begin{proof}[Proof of \cref{lem:x1-upper-bound}]
    For each of the $n$ vertices, it is isolated with probability $(1 - p)^{n - 1}$.
    It follows from linearity of expectation  
    \begin{align*}
        \E{x_1}
            &= n \cdot (1 - p)^{n - 1} 
            \le n \cdot e^{-p \cdot (n - 1)} 
            \le n \cdot e^{ - 2 \cdot \ln n } 
            = \frac{1}{n}.
    \end{align*}
\end{proof}

\begin{fact}[\cite{Blum_Hopcroft_Kannan_2020}]
    The expected number of connected components of size $k$ in a Erdős–Rényi model $\cG(n, p)$, denoted by $\E{x_k}$, is bounded by  
    \begin{equation*}\label{eq:numconcomp}
        \E{x_k} \le \binom{n}{k} \cdot k^{k - 2} \cdot p^{k - 1} \cdot (1 - p)^{k (n - k)}.
    \end{equation*}
\end{fact}

\begin{corollary}
    \label{lem:small-upper-bound}
    Assuming that $n \ge 1700$.
    If $p = \frac{c + \ln n}{n}$ for $c \ge 2 \ln (2 \ln n) + 2$, then 
    \begin{equation}
        \sum_{k = 2}^{n / 2} \E{x_k} \le \frac{2}{n}.
    \end{equation}
\end{corollary}

The proof of \cref{lem:small-upper-bound} is deferred to the end of this section.  
We now proceed to prove \cref{lem:erconnectivity}.  

If the graph is not connected, there must be a connected components of size from $\{1, 2, \dots, n - 1\}$. 
Therefore it's sufficient to show that $\sum_{k = 1}^{n - 1}\E{x_k} \le \frac{6}{n}$. 
Note that in the random graph, there can be at most one connected components of size in $(n/2, n)$. 
Furthermore if there exists such a connected component of size in $(n/2, n)$, there must be at least one connected components of size in $[1, n/2]$, therefore for any random graph
\begin{align}
    \sum_{k = 1}^{n/2}x_k \ge \sum_{k > n/2}^{n - 1}x_k ~.
\end{align}
From \cref{lem:x1-upper-bound} and $\cref{lem:erconnectivity}$, we have
\begin{align}
    \sum_{i = 1}^{n/2}\E{x_k} \le \frac{3}{n} ~.
\end{align}
Therefore, 
\begin{align}
    \sum_{k = 1}^{n - 1}\E{x_k} = \sum_{k = 1}^{n/2}\E{x_k} + \sum_{k > n/2}^{n - 1}x_k \le 2\sum_{k = 1}^{n/2}\E{x_k} \le \frac{6}{n},
\end{align}
which finishes the proof of \cref{lem:erconnectivity}.
\end{proof}

\begin{proof}[Proof of \cref{lem:small-upper-bound}]
    Since $\binom{n}{k} \le \PAREN{\frac{en}{k}}^k, 1 - x \le e^{-x}$, if $p = \frac{c + \ln n}{n}$, then
    \begin{align*}
        \E{x_k}
            &\le \binom{n}{k} \cdot k^{k - 2} \cdot p^{k - 1} \cdot (1 - p)^{k (n - k)} \\
            &\le \PAREN{\frac{en}{k}}^k\cdot k^{k - 2} \cdot p^{k - 1} \cdot (1 - p)^{k (n - k)} \\
            &\le \frac{1}{k^2} \cdot e^k \cdot n^k \cdot \PAREN{\frac{c + \ln n}{n}}^{k - 1} \cdot e^{- \frac{c + \ln n}{n} \cdot k (n - k)} \\
            &= \frac{1}{k^2} \cdot \PAREN{\frac{c + \ln n}{n}}^{k - 1} \cdot e^{- \frac{k (n - k) - k n}{n} \cdot \ln n - \frac{c \cdot k (n - k) - k n}{n}  } \\
            &= \frac{1}{k^2} \cdot \PAREN{\frac{c + \ln n}{n}}^{k - 1} \cdot e^{\frac{k^2}{n} \cdot \ln n - \frac{k (c n - c k - n) }{n}  }.
    \end{align*} 
    When $k \le n / 2$, we have
    \begin{equation*}
        cn - ck - n 
            \ge cn - cn / 2 - n 
            = (c / 2 - 1) n.
    \end{equation*}
    Therefore, 
    \begin{align*}
        \PAREN{c + \ln n}^{k - 1} \cdot e^{- \frac{k (c n - c k - n) }{n} }
            &\le e^{
                - k \cdot \PAREN{ \frac{c}{2} - 1 - \ln (c + \ln n) }
            }.
    \end{align*}   
    The function $y = c / 2 - 1 - \ln (c + \ln n)$ is minimized when $c = 2 \ln (2 \ln n) + 2$, which gives 
    \begin{align*}
        &\ln (2 \ln n) + 1 - 1 - \ln (2 \ln (2 \ln n) + 2 + \ln n) \\
        &= \ln (2 \ln n) - \ln (2 \ln (2 \ln n) + 2 + \ln n) 
        \ge 0,
    \end{align*}
    where the last inequality holds if $n \ge 1700$. 
    It follows that 
    \begin{align*}
        \E{x_k}
            \le \frac{1}{k^2} \cdot \PAREN{\frac{1}{n}}^{k - 1} \cdot e^{\frac{k^2}{n} \cdot \ln n } 
            = \frac{1}{n} \cdot e^{- 2 \ln k - (k - 2) \cdot \ln n + \frac{k^2}{n} \cdot \ln n }
    \end{align*} 
    The function $g(k) \doteq - 2 \ln k - (k - 2) \cdot \ln n + \frac{k^2}{n} \cdot \ln n$ is convex respect to $k$, therefore obtaining maximum either at $k = 2$ or $k = n / 2$. 
    We have 
    \begin{align*}
        g(2) 
            &= - 2 \ln 2 + \frac{4 \ln n}{n} \le 0, \, &\text{ if } n \ge 4, \\
        g(3) &= -2 \ln 3  - \ln n + \frac{9 \ln n}{n} \le -\ln n, \, &\text{ if } n \ge 9, \\
        g(n / 2) &= - 2 \ln \frac{n}{2} - \PAREN{\frac{n}{2} - 2} \cdot \ln n + \frac{n}{4} \cdot \ln n \\
        &= 2 \ln 2 - \frac{n}{4} \cdot \ln n \le - \ln n, \, &\text{ if } n \ge 9.
    \end{align*}
    It follows that 
    \begin{align*}
        \E{x_2} \le \frac{1}{n}, \text{ and } 
        \E{x_k} \le \frac{1}{n^2}, \forall 3 \le k \le n / 2. 
    \end{align*}
    It concludes that $\sum_{k = 2}^{n / 2} \E{x_k} \le \frac{2}{n}.$
\end{proof}

\hfill $\square$

\vspace{10mm}
\section{Proof of \cref{lemma: maximum of subgaussin}}
\label{proof: maximum of subgaussin}
\begin{proof}
    Since $w_e$ is a sum of $\numSample$ independent random variables between $[0, 1]$, and has expectation $\numSample \cdot P_e$, by Hoeffding's inequality, it holds that 
    \begin{align}
        \P{ |w_e - \numSample \cdot P_e| \ge t }
            \le 2 \cdot \exp \PAREN{
                - \frac{2t^2}{\numSample}
            }, 
        \quad 
        \forall t \ge 0.
    \end{align}
    By union bound, it holds that for all $t \ge 0$, $s \ge 1$ and $n \ge 100$ 
    \begin{align*}
        \P{ \max_{e \in E} |w_e - \numSample \cdot P_e | \ge t + \sqrt{\numSample \cdot \ln n} }
            &\le 2 \cdot \frac{n (n - 1)}{2} \cdot \exp \PAREN{
                -2 \cdot \PAREN{ t + \sqrt{\numSample \cdot \ln n}  }^2 / \numSample
            } \\ 
            &\le n^2 \cdot \exp \PAREN{
                -2 \cdot t^2 / \numSample
            } \cdot \exp \PAREN{
                -2 \cdot \numSample \cdot (\ln n) / \numSample
            } \\
            &= \exp \PAREN{
                -2 \cdot t^2 / \numSample
            }.
    \end{align*}
    Therefore, 
    \begin{align*}
        \E{ \max_{e \in E} |w_e - \numSample P_e | }
            &= \int_0^{\sqrt{\numSample \ln n}} \P{ \max_{e \in E} |w_e - \numSample \cdot P_e | \ge t } \, d t
            + \int_{\sqrt{\numSample \ln n}}^{\infty} \P{ \max_{e \in E} |w_e - \numSample \cdot P_e | \ge t } \, d t \\
            &\le \sqrt{\numSample \cdot \ln n}
            + \int_0^{\infty} \P{ \max_{e \in E} |w_e - \numSample \cdot P_e | \ge t + \sqrt{\numSample \cdot \ln n} } \, d t \\
            &\le \sqrt{\numSample \cdot \ln n} + \int_0^\infty \exp \PAREN{
                -2 \cdot t^2 / \numSample
            } \, d t \\
            &= \sqrt{\numSample \cdot \ln n} + \frac{1}{2} \cdot \int_{-\infty}^\infty \exp \PAREN{
                - \frac{t^2}{ 2 \cdot (\numSample / 4) }
            } \, d t \\
            &= \sqrt{\numSample \cdot \ln n} + \frac{\sqrt{2 \cdot \pi \cdot (\numSample / 4)}}{2} \\
            &< \sqrt{3\numSample\ln n}.
    \end{align*}
\end{proof}

\newpage
\section{Experiments}\label{appx:experiments}

We present two experiments that support our theoretical claims.

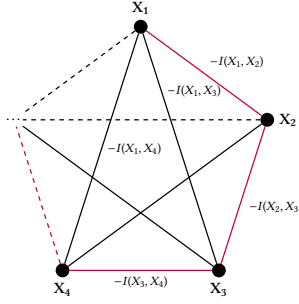
\begin{figure}
    \centering
    \resizebox{0.3\columnwidth}{!}
    {     \begin{tikzpicture}
    [edge/.style = {draw, thick}, edge pink/.style = {draw=purple, thick }, 
    vertex/.style args = {#1 #2}{circle, draw, thick, inner sep=0pt, fill=black, minimum size=8pt, label=#1:#2},
    dotsvertex/.style = {thick, inner sep=0pt, minimum size=8pt} ]
    \node (a) [vertex=above $\mathbf{X_1}$] at (90:3cm) {};
    \node (b) [vertex=right $\mathbf{X_2}$] at (18:3cm) {};
    \node (c) [vertex=below $\mathbf{X_3}$] at (306:3cm) {};
    \node (d) [vertex=below $\mathbf{X_4}$] at (234:3cm) {};
    \node (e) [dotsvertex] at (162:3cm) {\dots};
    
    \draw[edge pink] (a) -- (b) node[midway, above right] {\footnotesize{$-I(X_1, X_2)$}};
    \draw[edge] (a) -- (d) node[midway, right] {\footnotesize{$-I(X_1, X_4)$}};
    \draw[edge] (a) -- (c) node[near start, right] {\footnotesize{$-I(X_1, X_3)$}};
    \draw[edge] (a) -- (e) [dashed] node[midway, above left] {};
    \draw[edge pink] (b) -- (c) node[midway, below right] {\footnotesize{$-I(X_2, X_3)$}};
   \draw[edge] (b) -- (d) node[midway, below] {};
    \draw[edge] (b) -- (e) [dashed] node[midway, above] {};
    \draw[edge pink] (c) -- (d) node[midway, below] {\footnotesize{$-I(X_3, X_4)$}};
    \draw[edge] (c) -- (e) node[midway, left] {};
    \draw[edge pink] (d) -- (e) [dashed] node[midway, below left] {};
    \end{tikzpicture}   }
    \caption{An extract of the complete graph encoding the mutual information between the random variables $X_1, ..., X_n$ described in \cref{def:process} and used in \cref{sec:mi}.
    The weights encode the negated mutual information corresponding to the described process. 
    The mst is formed by the vertices on the path $P(X_1,X_2, \dots)$.
    In our experiment with $n=1000$ vertices and the flip probability $p=0.05$, we have $-I(X_1, X_2) = I(X_2, X_3) = \dots \approx -0.7136, -I(X_1, X_3) = I(X_2, X_4) = \dots \approx -0.5471$ 
    and $-I(X_1, X_4) \dots \approx -04277$.
    }
    \label{fig:mi-graph}
\end{figure}

\paragraph{Mutual Information}\label{sec:mi}
One natural setting is finding the Chow-Liu tree \cite{chowliu1968}, which is the mst on the graph encoding the negated mutual information matrix on all pairwise attributes.\footnote{Note that in the Chow-Liu setting, we want to find the maximum spanning tree, which is the same as the mst on the negated weights.}
Let $X_1, \dots, X_n\in \{0, 1\}$ be random bits.
We draw $X_1 \sim Ber(1/2)$, and then recursively $X_{i}$ by flipping the bit $X_{i-1}$ with probability $0 < p < 1/2$. 
This simulates a natural scenario where there is mutual information between $X_i$ and $X_{i-1}$ controlled by $p$.
The mst is formed by the edges on the path $P = (X_1, ... X_n)$ (visualized in  \cref{fig:mi-graph}).
A potential underlying dataset of size $d$ has the sensitivity of mutual information, which is $\Delta_{mi} = \log_2(d)/d$.
As later shown in \cref{appx:mutual-information}, for this process the pairwise mutual information between all $X_i$ and $X_j$ define a complete graph $G = (V, E)$ with weights
\[
\forall i,j  \in [n], i\neq j: w_{ij} = \dfrac{1}{2}\left(p_1 \log_2 \left(p_1\right) + p_2 \log_2\left(p_2\right)\right)
\]
where $p_1 = 1+(1-2p)^k$ and $p_2 =  1-\left(1-2p\right)^k$.

\begin{figure}[ht]
  \centering
    \subfloat[\centering Mutual Information Graph \protect\linebreak (Experiment 1)]{{\includegraphics[width=6.5cm]{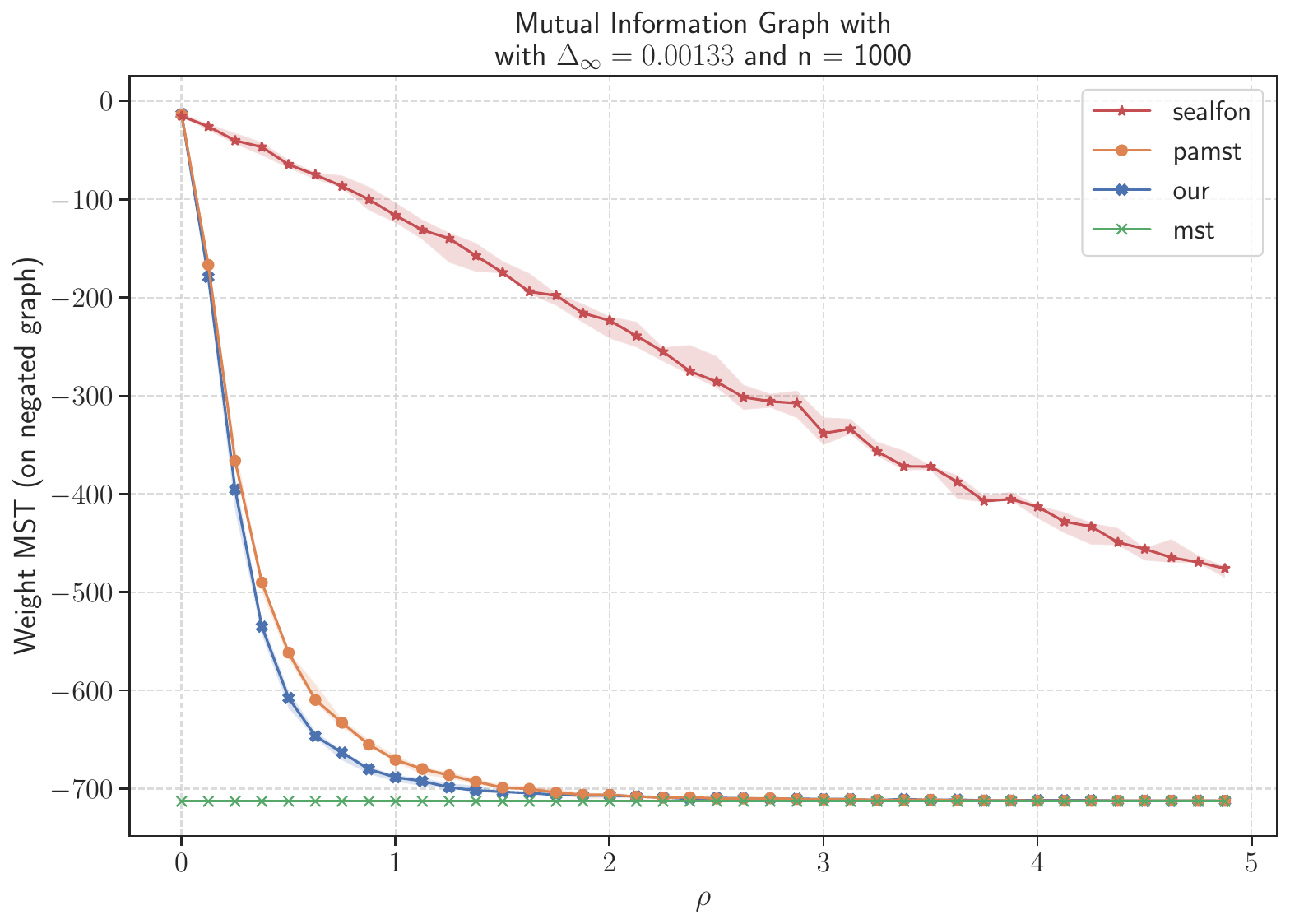} }}%
    \qquad
  \subfloat[\centering Effect of density on a random graph\protect\linebreak (Experiment 2)]{{\includegraphics[width=6.5cm]{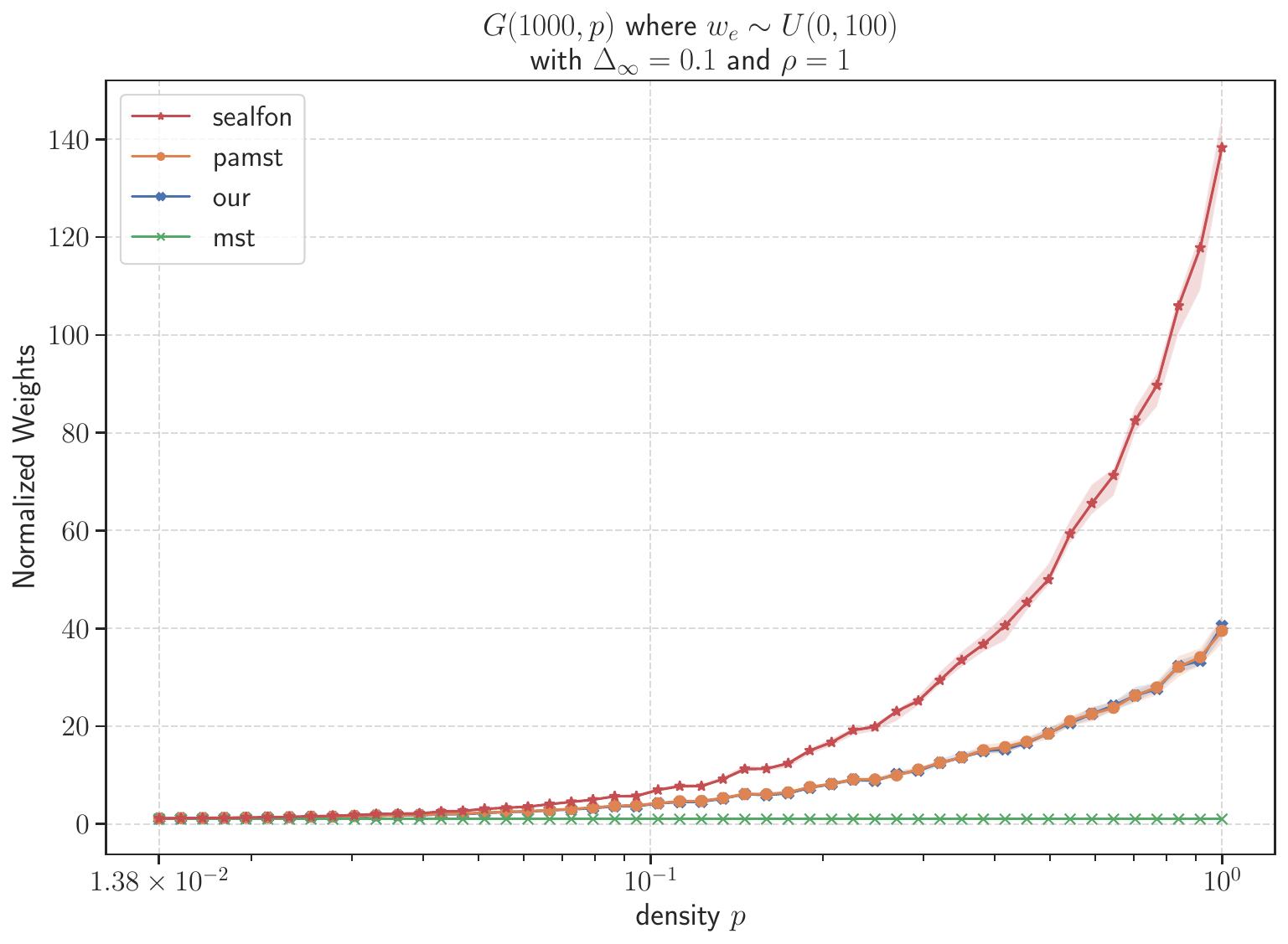} }}%
  \caption{The results of our experiment. \textbf{a)} Shows the instance instance described in Experiment 1. 
  Note that we have to negate all weights to find the maximum spanning tree on the mutual information graph. \textbf{b)} Shows the impact of the graph's density on random graphs with $n=1000$ vertices for a fixed privacy level $\rho = 1$: 
  The figure shows the ratio between the real mst and the private one, where each edge weight is uniformly drawn from the interval $[0,100]$.
  Because the noise scale of Sealfon's input perturbation scales with the number of edges in the graph, we see a larger gap for denser graphs.
  Each data point shows the median of ten runs. 
  } \label{fig:results}
\end{figure}

In the experiment, we used $n=1000$ with flip probability of $p = 0.05$.
We set the underlying dataset size to $d = 10^5$, thus $\Delta_\infty = \Delta_{mi} \approx 0.00133$. 
The results are shown in \cref{fig:results}-\textbf{a)}.
We can see that the error of our approach closely resembles PAMST, outperforming the input privatization approach.

\paragraph{The Effect of the density}\label{sec:density}

The second experiment follows the setup proposed in \cite{pinot_2018_ma}.
We construct a random graph using the Erdős–Rényi model $\cG(n, p)$ and include an edge $e$ with probability $p$.
We draw the weights $w_e \sim U(0, 100)$ uniformly.
The experiments shown in \cref{fig:results}-\textbf{b)} were run on graphs of size $n = 1000$ using $\Delta_\infty = 0.1$.
As shown, \emph{input privatization} does significantly worse as the graph's density increases, whereas our approach resembles the output distribution of PAMST.

\subsection{Proofs for the empirical evaluation}\label{appx:mutual-information}

Assume $0<p<\frac{1}{2}$.
The mutual information $I(X; Y)$ between two discrete random variables $X$ and $Y$ quantifies the amount of information that $X$ contains about $Y$ (or vice versa).

\begin{definition}[Mutual Information \cite{ash1990information}]
\begin{align}
    I(X; Y) = \sum_{x \in \mathcal{X}} \sum_{y \in \mathcal{Y}} \Pr[X = x, Y = y] \log_2 \left( \frac{\Pr[X = x, Y = y]}{\Pr[X = x] \Pr[Y = y]} \right),
\end{align}
\end{definition}

\noindent We now prove the validity of the weight function given in \cref{sec:mi}.

\begin{definition}\label{def:process}
We define the random variables $X_1,\dots,X_n \in \{0,1\}$ recursively by setting 
\begin{align*}
X_1 \sim Ber\left(\dfrac{1}{2}\right)  \text{~~~and~~~} & X_i= \begin{cases}
       1 - X_{i-1} & \text{with probability~} p \\
       X &\text{else}
    \end{cases}
\end{align*}
\end{definition}
By definition, $\Pr[X_i = 0] = \Pr[X_i] = 1 = \frac{1}{2}$. 
Note that mutual information is minimized for $p\approx \frac{1}{2}$ and note that for all $X_i: \Pr[X_i = 1] = \Pr[X_i = 0] = \frac{1}{2}$. 
We need another short lemma for the parity of a binomial random variable:

\begin{lemma}[Parity of Binomials.]
Denote $Z\sim \cB(n,p)$, then
\begin{align}
    Pr[\text{Z is even}] = 1- \Pr[\text{Z is odd}] = \frac{1}{2}+ \frac{1}{2}(1-2p)^n
\end{align}
\end{lemma}
\begin{proof} \begin{align*}
\Pr\limits_{Z \sim \cB(n, p)}[\text{Z is even}] &= \sum\limits_{k = 0}^{n}\frac{(-1)^k+1}{2} \Pr[Z = k]\\
&= \frac{1}{2} \left(\sum\limits_{k=0}^n((-1)^k \Pr[Z = k] + \sum\limits_{k=0}^n \Pr[Z = k]\right) \\
&= \frac{1}{2} \left(\sum\limits_{k=0}^n\left((-1)^k \Pr[Z = k]\right) + 1\right) \\
&= \frac{1}{2} \left(\sum\limits_{k=0}^n\left((-1)^k \binom{n}{k}p^k (1-p)^{n-k}\right) + 1\right) \\
&= \frac{1}{2} \left(\sum\limits_{k=0}^n\left(\binom{n}{k}(-p)^k (1-p)^{n-k}\right) + 1\right) \\
&=\frac{1}{2}(1-2p)^n+\frac{1}{2} \\
\end{align*}
where the last line follows from the binomial theorem.
\end{proof}
We can now directly compute the mutual information in this process.
\begin{claim}
Assume $0<p<\frac{1}{2}$, the mutual information score between $X_i$ and $X_j$ that are $k = |i-j|$ steps apart can directly be computed by 
\begin{align}
    I(X_i; X_j)=\left(\frac{1}{2}+\frac{1}{2}(1-2p)^k\right)\log_2 \left(1+(1-2p)^k\right) + \left(\frac{1}{2}-\frac{1}{2}\left(1-2p\right)^k\right)\log_2\left(1-(1-2p)^k)\right)
\end{align}
\end{claim}
\begin{proof}

We shortly write $p_{00}(k) =\Pr[X_i = 0, X_j = 0]$ and $p_0 = \Pr[X_i = 0]$ (resp $p_{11}(k)$, $p_{10}(k)$ and $p_{01}(k)$, $p_{1}$).
Furthermore, denote $p_{even}(k) = \Pr\limits_{Z \sim \cB(k,p)}[\text{Z is even}]$ and $p_{odd}(k)$ as proven above.
Note that we can compute these probabilities:
\begin{align}
    p_{00}(k) &= \Pr[X_i = 0| X_j = 0]\cdot\Pr[X_j = 0] = p_{even}(k) \cdot \frac{1}{2} = \frac{1}{4}+\dfrac{1}{4}\cdot(1-2p)^k\\
    p_{10}(k) &= \Pr[X_i = 1|X_j = 0]\cdot\Pr[X_j = 0] = p_{odd}(k) \cdot \frac{1}{2} = \frac{1}{4}-\frac{1}{4}(1-2p)^k
\end{align}
We trivially get $p_{11}(k)=  p_{00}(k)$ and $p_{01}(k) = p_{10}(k)$ by symmetry.
Then, we can compute the mutual information directly. 
\begin{align*}
I(X_i; X_j) &= p_{00}(k) \log_2 \frac{p_{00}(k)}{p_0p_0} + p_{01}(k) \log_2 \frac{p_{01}(k)}{p_0p_1} + p_{10}(k) \log_2 \frac{p_{10}(k)}{p_{1}p_{0}} + p_{11}(k) \log_2 \frac{p_{11}(k)}{p_1p_1} \\
&= p_{00}(k)\log_2 (4p_{00}(k)) + p_{01}(k) \log_2 (4 p_{01}(k)) + p_{10}(k) \log_2 (4 p_{10}(k)) + p_{11}(k) \log_2 (4 p_{11}(k))\\
&= 2p_{00}(k)\log_2 (4p_{00}(k)) + 2p_{01}(k) \log_2 (4 p_{01}(k)) \\
&=\left(\frac{1}{2}+\frac{1}{2}(1-2p)^k\right)\log_2 \left(1+(1-2p)^k\right) + \left(\frac{1}{2}-\frac{1}{2}\left(1-2p\right)^k\right)\log_2\left(1-(1-2p)^k)\right)
\end{align*}
\end{proof}

\newpage
\section{Table of symbols}
\begin{table*}[h]
\centering
\resizebox{0.80\columnwidth}{!}{%
    \begin{tabular}{cl}
    {\bf Symbol} & {\bf Description} \\
    PPSACR &  Probability Proportional to Sizes with Adaptive Candidate Removal\\
    $MST$ & Minimum Spanning Tree \\[0.65em]
    $G = (V, E, \vec{W})$ & Graph  \\
    $\vec{W}\in \R^{|E|}; w_e $ & Weights of $G$ \\
    $n$ & Number of vertices \\
    $m$ & Number of edges \\
    $T\subseteq E$ & A tree in $G$ \\
    $\cT(G)$ & Set of all spanning trees \\[0.65em]
    $\Delta_1$, $\Delta_2$, $\Delta_\infty$ & Sensitivity parameters \\
    $W \sim W'$ & Neighboring weights\\
    $\rho, \epsilon, \delta$ & Privacy parameters \\[0.65em]
    $\beta$ & High probability bound, also used in Beta dist \\
    $I(X; Y)$ & Mutual Information between $X$ and $Y$\\
    $\cB(k, p)$ & Binomial Distribution\\
    $\ExpNoise{\lambda}$ & Exponential Distribution\\
    $\GumbelNoise{b}$ & Gumbel Distribution\\
    $\cG(n, p)$ & Erdos-Renyi graphs \\[0.65em]
    $\tilde{\mathcal{O}}(f)$, $\tilde{\Omega}(f)$, $\tilde{\Theta}(f)$ & Hides log factors\\
    \end{tabular} }\\
\caption{Symbols used throughout the paper}
\label{tab:symbols}
\end{table*}

\end{document}